%% file: main.tex
\documentclass[CERN,manyauthors]{cernphprep}
\usepackage{geometry}
\usepackage[comma,square,numbers,sort&compress]{natbib}
\usepackage[pdftitle={Study of hard and electromagnetic processes at SPS energy},
        pdfauthor={NA60+ Collaboration},
        pdfsubject={},
        bookmarksopen=false]{hyperref}
\usepackage{lineno}
\usepackage{bm}
\usepackage{xcolor}
\usepackage{textcomp}
\usepackage{parskip}
\usepackage{enumitem}
\setlist{nolistsep}

\input{definitions}

\begin{document}%

\begin{titlepage}
\title{Study of hard and electromagnetic processes at CERN-SPS energies:\\ an investigation of the high-$\mu_{\mathbf{B}}$ region of the QCD phase diagram with NA60+}
\ShortTitle{Study of hard and electromagnetic processes at CERN-SPS energies}

\Collaboration{NA60+ Collaboration\thanks{See \hyperref[app:collab]{Appendix} for the list of collaboration members}}
\ShortAuthor{NA60+ Collaboration}
\bigskip
\begin{abstract}
\medskip
The exploration of the phase diagram of Quantum ChromoDynamics (QCD) is carried out by studying ultrarelativistic heavy-ion collisions. The energy range covered by the CERN SPS ($\sqrt{s_{\rm \scriptscriptstyle{NN}}} \sim 6\text{--}17$~GeV) is ideal for the investigation of the region of the phase diagram corresponding to finite baryochemical potential ($\mu_{\rm B}$), and has been little explored up to now. We propose in this document a new experiment, NA60+, that would address several observables which are fundamental for the understanding of the phase transition from hadronic matter towards a Quark--Gluon Plasma (QGP) at SPS energies. In particular, we propose to study, as a function of the collision energy, the production of thermal dimuons from the created system, from which one would obtain a caloric curve of the QCD phase diagram that is sensitive to the order of the phase transition. In addition, the measurement of a $\rho\text{--}{\rm a}_1$ mixing contribution would provide conclusive insights into the restoration of the chiral symmetry of QCD.
In parallel, studies of heavy quark and quarkonium production would also be carried out, addressing the measurement of transport properties of the QGP and the investigation of the onset of the deconfinement transition. The document also defines an experimental set-up which couples a vertex telescope based on monolithic active pixel sensors (MAPS) to a muon spectrometer with tracking (GEM) and triggering (RPC) detectors within a large acceptance toroidal magnet.
Results of physics performance studies for most observables accessible to NA60+ are discussed, showing that the results of the experiment would lead to a significant advance of our understanding of strong interaction physics.
The document has been submitted as an input to the European Particle Physics Strategy Update 2018--2020 
(\url{http://europeanstrategyupdate.web.cern.ch/}).
\end{abstract}
\vfill
{\centering\large\today\par}
\end{titlepage}

\setcounter{page}{1}
\input{Introduction}
\input{Observables}
\input{Requirements}
\input{Experiment}
\input{PhysicsPerformances}
\input{Competition}
\input{Summary}
%

\newenvironment{acknowledgement}{\relax}{\relax}
\begin{acknowledgement}
\section*{Acknowledgements}
We would like to thank F.~Bergsma and P.\ A.\ G.\ Giudici (both CERN) for their contribution on the study of the toroidal magnet.
\end{acknowledgement}
%

\bibliographystyle{utphys}
\bibliography{main}

\newpage
\appendix
\section*{Appendix: NA60+ Collaboration}
\label{app:collab}
\bigskip
\input{authorlist-preprint.tex}
\end{document}

%% file: definitions.tex
\newcommand{\pp}{\ensuremath{\rm pp}\xspace}

\newcommand{\pA}{\ensuremath{\text{pA}}\xspace} 
\newcommand{\pBe}{\ensuremath{\text{p--Be}}\xspace}
\newcommand{\pPb}{\ensuremath{\text{p--Pb}}\xspace}
\newcommand{\pU}{\ensuremath{\text{p--U}}\xspace}

\newcommand{\AuAu}{\ensuremath{\text{Au--Au}}\xspace}

\newcommand{\PbPb}{\ensuremath{\text{Pb--Pb}}\xspace}
\newcommand{\InIn}{\ensuremath{\text{In--In}}\xspace}

\newcommand{\sqrtsNN}{\ensuremath{\sqrt{s_{\mathrm{\scriptscriptstyle NN}}}}\xspace}

\newcommand{\Elab}{\ensuremath{E_{\rm lab}}\xspace}

\newcommand{\aone}{\ensuremath{\mathrm{a}_1}\xspace}
\newcommand{\jpsi}{\ensuremath{\mathrm{J}/\psi}\xspace}
\newcommand{\chic}{\ensuremath{\chi_{\rm c}}\xspace}

\newcommand{\psiP}{\ensuremath{\psi\text{(2S)}}\xspace}

\newcommand{\ccbar}{\ensuremath{\mathrm{c\overline{c}}}\xspace}

\newcommand{\Dmeson}[1]{\ensuremath{\mathrm{D}^{#1}}\xspace}

\newcommand{\Dzero}{\Dmeson{0}}
\newcommand{\Dplus}{\Dmeson{+}}

\newcommand{\Ds}{\ensuremath{\mathrm{D}^{+}_{s}}\xspace}

\newcommand{\lambdac}{\ensuremath{\Lambda_{\rm c}}\xspace}

\newcommand{\lambdacplus}{\ensuremath{\Lambda_{\rm c}^+}\xspace}

\newcommand{\ee}{\ensuremath{\mathrm{e}^+\mathrm{e}^-}\xspace}
\newcommand{\mumu}{\ensuremath{\mu^+\mu^-}\xspace}

\newcommand{\raa}{\ensuremath{R_{\mathrm{AA}}}\xspace}

\newcommand{\MeV}{\ensuremath{\text{~MeV}}\xspace}
\newcommand{\GeV}{\ensuremath{\text{~GeV}}\xspace}
\newcommand{\TeV}{\ensuremath{\text{~TeV}}\xspace}

\newcommand{\GeVc}{\ensuremath{\text{~GeV}/c}\xspace}

\newcommand{\eg}{e.\,g.\xspace}%
\newcommand{\ie}{i.\,e.\xspace}%

\newcommand{\dd}{\ensuremath{\mathrm{d}}}

\newcommand{\beq}{\begin{equation}}
\newcommand{\eeq}{\end{equation}}
\newcommand{\beqn}{\begin{eqnarray}}
\newcommand{\eeqn}{\end{eqnarray}}

\newcommand{\beqa}{\begin{eqnarray}}
\newcommand{\eeqa}{\end{eqnarray}}
\def\lsim{\raise0.3ex\hbox{$<$\kern-0.75em\raise-1.1ex\hbox{$\sim$}}}
\def\gsim{\raise0.3ex\hbox{$>$\kern-0.75em\raise-1.1ex\hbox{$\sim$}}}

\newcommand{\Npart}{\ensuremath{N_{\mathrm{part}}}\xspace}

\newcommand{\pt}{\ensuremath{p_{\mathrm{T}}}\xspace}

\newcommand{\muB}{\ensuremath{\mu_{\rm B}}\xspace}
\newcommand{\pythia}{{\sc Pythia}\xspace}
\newcommand{\powheg}{{\sc Powheg}\xspace}



%% file: Introduction.tex
\section{Introduction and QCD phase diagram}
\label{sec:introduction}

\begin{figure}[b]
\begin{center}
\includegraphics[width=0.49\linewidth]{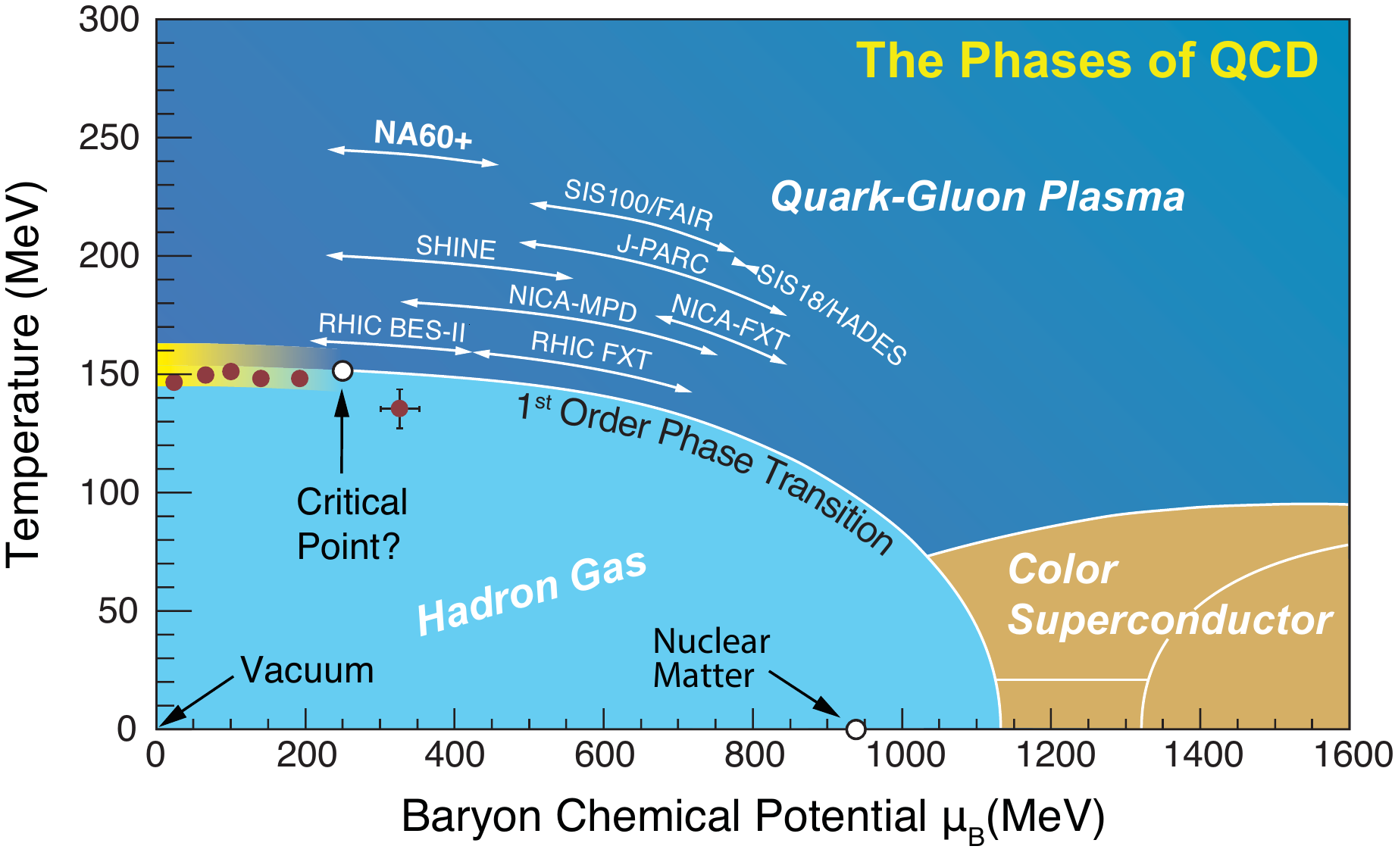}
\includegraphics[width=0.4\linewidth]{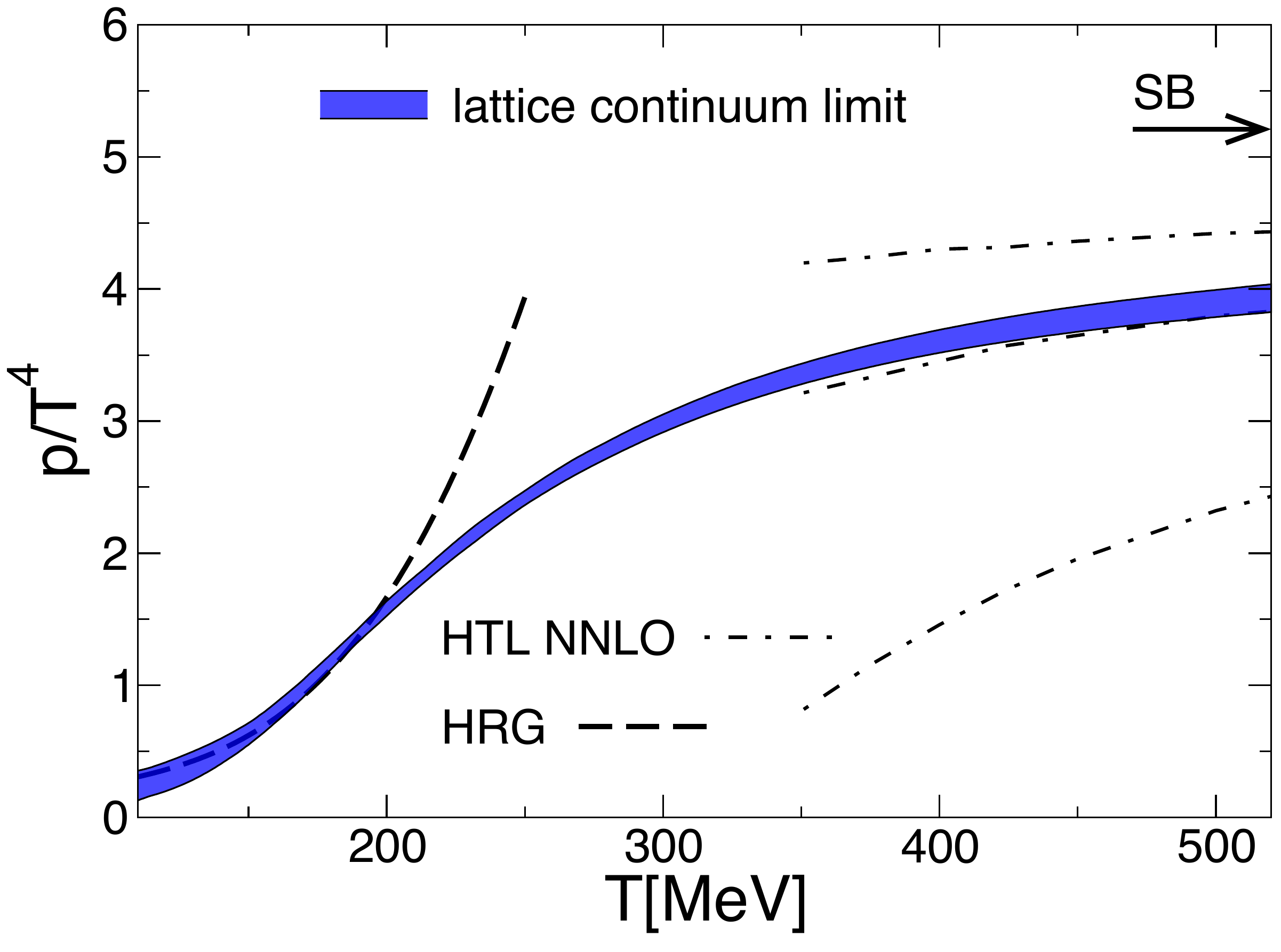}
\caption{The QCD phase diagram (left panel, courtesy of Thomas Ullrich) and the equation of state $P(T)/T^4$ in the limit of vanishing baryon density~\cite{Borsanyi:2013bia} measured on the lattice (right panel): the latter is characterized by a rise in the effective number of active degrees of freedom, indicating the cross-over transition to a QGP.}
\label{fig:QCDEOSmu}
\end{center}
\end{figure}

Quantum ChromoDynamics (QCD), the theory of the strong force, has a rich phase structure. On the one hand, asymptotic freedom allows it to completely define the degrees of freedom in terms of quarks and gluons, while on the other hand hadrons become the relevant degrees of freedom when confinement sets in. For a system in thermodynamic equilibrium, temperature ($T$) and baryochemical potential (\muB) are intensive parameters that are the same for any of its subsystems. It is convenient to describe the phase diagram, as shown in the left panel of Fig.~\ref{fig:QCDEOSmu}, in terms of these intensive parameters both from the perspective of lattice QCD in which the two are used as independent variables, and in the application of a grand-canonical ensemble in the analyses of heavy-ion experiments. Our quantitative understanding of the QCD phase diagram is largely restricted to the region of low \muB. For $\muB \sim 0$, lattice QCD provides quantitative results (see Fig.~\ref{fig:QCDEOSmu}-right): a fast increase of $\epsilon/T^4$ ($\epsilon$ = energy density) occurs around  a critical temperature $T_c\approx 155\MeV$. In this regime, the phase transition is a cross-over~\cite{Borsanyi:2013bia,Ratti:2018ksb}.

The extensive experimental campaigns conducted at the  CERN-SPS, BNL-RHIC and CERN-LHC accelerators have mostly explored so far this region of the phase diagram at low \muB, showing that a deconfined state of matter is produced in heavy-ion collisions at high energies, with properties consistent with the predictions of lattice QCD.
Present experiments at RHIC and SPS started in recent years to address features of the phase diagram structure at large \muB. For moderate temperatures and high baryon densities the existence of a first order phase transition with co-existence of a  mixed-phase was suggested. The first order transition line should end with a second order critical point. The existence of such a critical point and a first-order phase transition remain to be confirmed in experiments.
Evidence for a first order phase transition can be provided by the measurement of the caloric curve.  A direct measurement of such a curve for the phase transition between a hadron gas and a QGP can be based on precise temperature measurements as a function of the centre of mass energy. 

Chiral symmetry is spontaneously broken in the hadronic world due to the formation of quark and gluon condensates. This leads to the generation of the hadron mass splittings in the light hadron spectrum of ${\sim}0.5\GeV$ for the chiral partners $\pi\text{--}\sigma$ and $\rho\text{--}\aone$. Lattice QCD calculations for $\muB = 0$ show that at the hadron-QGP phase transition boundary the chiral condensate steeply decreases around $T_c$, indicating the approach to chiral restoration~\cite{Borsanyi:2010bp}.

Precise measurements of heavy quarkonium states can be used to study the suppression of such states due to colour screening which is considered a key signature of a deconfined state. On the other hand, open charm measurements allow us to extract fundamental transport coefficients from the QGP as well as improve our understanding of hadronization mechanisms.

We propose a new experiment, NA60+, as a follow-up experiment to NA60~\cite{Baldit:2000cq,Arnaldi:2008er} to address each of these sets of observables through very high precision measurements in \PbPb collisions in the finite \muB region of the phase diagram where such measurements either have not been done, or have not yet reached a level of precision that would allow for meaningful constraints on theoretical models. In the following section~\ref{sec:observables}, we discuss the NA60+ physics  plans and how they will address each of the aforementioned observables through high precision dimuon and hadronic measurements. In the sections~\ref{sec:requirements} and~\ref{sec:na60plus} we detail the requirements to accomplish this programme in terms of collision energies,  integrated luminosities that allow for meaningful measurements, and proposed detector technologies, respectively. Based on those requirements, section~\ref{sec:physics_performances} provides projections for the physics performance for each of the observables discussed in section~\ref{sec:observables}. This is followed by a brief overview of other initiatives in terms of present and future detectors/facilities in section~\ref{sec:competition} and our conclusions in strong favour of the NA60+ project in section~\ref{sec:summary}.

%% file: Observables.tex
\section{Observables/Measurements}
\label{sec:observables}
\subsection{Thermal radiation}
\label{subsec:obsthermalradiation}
\subsubsection{Chiral symmetry restoration: measurement of \texorpdfstring{$\mathbf{\rho\text{--}a_1}$}{rho--a1} chiral mixing}
\label{subsec:chiralsymmetry}

A long-standing question is how chiral symmetry restoration manifests itself in the hadron spectrum, \ie\ what are its observable consequences.
The NA60 experiment precisely measured the $\rho$ spectral properties in \InIn collisions at $\sqrtsNN = 17.3\GeV$~\cite{Arnaldi:2006jq,Arnaldi:2008er,Specht:2010xu} to be in agreement with a microscopic many-body model predicting a very strong broadening with essentially no mass shift~\cite{Rapp:1999us,vanHees:2007th}. To understand if the $\rho$ {\it melting} is a signal of the restoration of chiral symmetry, and more in general to understand the changes in the hadron spectrum, a study of the spectral properties of its chiral partner \aone is fundamental. We present here our proposal to investigate the $\rho\text{--}\aone$ chiral mixing towards chiral symmetry restoration.

There is no direct coupling of axial states to the dilepton channel, so that  in vacuum the $\ee\to\text{hadrons}$ cross section has a dip in the \aone mass range~\cite{PhysRevD.98.030001}. However, in the medium, for dilepton masses above 1\GeV, 4 pion (and higher) states dominate the free electromagnetic correlator. Model-independent predictions of chiral symmetry show that to leading order in temperature one has a pion-induced mixing of  vector ($V$) and axial-vector ($A$) correlators~\cite{DEY1990620}:
$\Pi_V(T) = (1-\epsilon)\Pi^{vac}_V + \epsilon \Pi^{vac}_A$ where $\epsilon$ is the mixing parameter ($\epsilon = T^2/6f^2_{\pi}$, $f_{\pi} = 93\MeV$, and $\epsilon \sim 0.5$ at chiral symmetry restoration). The admixture of the \aone resonance, via the axial-vector correlator, thus entails an enhancement of the dilepton rate for $M\sim1\text{--}1.4\GeV$. A precision measurement of this enhancement for $M>1\GeV$ sheds light on the $\rho\text{--}\aone$ chiral mixing and the approach to chiral symmetry restoration.

To optimize the dilepton signal of interest, the lifetime of matter around $T_c$ should be maximized relative to other contributions: starting out in a (soft) mixed phase is a key point to enhance the effects related to chiral restoration. Thus the collision energy should be large enough to produce this matter, but small enough not to overshoot it producing partonic matter. This is expected to occur at collision energies somewhat smaller than $\sqrtsNN=17\GeV$, close to the onset of deconfinement, where the QGP yield is negligible. The optimal energy range of interest can be then established by the medium temperature measurement, as detailed in section \ref{subsec:caloriccurve}.

A calculation of the dilepton thermal yield with no chiral mixing and full chiral mixing at $\sqrtsNN = 8.8\GeV$ ($\epsilon = 1/2$) has been performed within the framework of~\cite{RAPP2016586}. The chiral mixing will fill the dip in the mass range 1--1.4\GeV with a yield increase of ${\sim}30\%$.  

\subsubsection{Hadron--QGP phase transition: measurement of the strongly interacting matter caloric curve at high \texorpdfstring{$\mu_{\mathbf{\textit B}}$}{muB}}
\label{subsec:caloriccurve}

The measurement of a caloric curve provided evidence for a first order phase transition from the liquid self-bound nuclear ground state to a gas of free nucleons~\cite{PhysRevLett.75.1040}. We present here the method to perform the first measurement of a caloric curve for the phase transition between the hadron gas and the QGP. The temperature measurement, performed as a function of collision energy, is based on a precise {\it dimuon} thermometer.

For dilepton masses above 1.5\GeV, the continuum of overlapping resonances leads to a flattened spectral density corresponding to a simpler description in terms of quarks and gluons (hadron-parton duality). Here medium effects on the electromagnetic spectral function ($\Pi_{\rm EM}$) are parametrically small, of the order of $(T/M)^2$, providing a stable thermometer. With $\Pi_{\rm EM}\propto M^2$, and in non-relativistic approximation, one has $\dd N/\dd M \propto M^{3/2}\exp(-M/T_s)$~\cite{RAPP2016586}. Since  mass is by construction a Lorentz-invariant, the mass spectrum is immune to any motion of the emitting sources, unlike transverse momentum (\pt) spectra.  The parameter $T_s$ in the spectral shape of the mass spectrum is therefore a space-time average of the thermal temperature $T$ over the  fireball evolution. The choice of the mass window, 1.5--2.5\GeV, implies $T\ll M$ and thus strongly enhances the sensitivity to the early high-$T$ phases of the evolution.
This method has been exploited by NA60 to measure the medium temperature in In-In collisions at $\sqrtsNN = 17.3\GeV$. The fit of the mass spectrum gives $T_s = 205\pm12\MeV$~\cite{Arnaldi:2008er,Specht:2010xu}. This is above $T_c$, thus showing that the QGP is produced at this collision energy.

The evolution of the initial temperature and $T_s$ vs collision energy has been studied theoretically in~\cite{RAPP2016586} in the interval $\sqrtsNN = 6\text{--}200\GeV$. However, in this calculation a cross-over transition is assumed. The average temperature $T_s$ from the mass fit is about 30\% below the corresponding initial one at $\sqrtsNN = 200\GeV$. The two temperatures are both close to the critical temperature below $\sqrtsNN = 10\GeV$, with their gap reducing to less than 15\%. This is largely due to the \mbox{(pseudo-)}latent heat in the transition which needs to be burned off in the expansion/cooling. The collision energy range below $\sqrtsNN=10\GeV$ thus appears to be essential to map out the phase-transition regime at high \muB, with the possible discovery of a plateau in the caloric curve built with dilepton slopes $T_s$. The experimental programme of NA60+ thus proposes to perform an energy scan in the interval $\sqrtsNN=6\text{--}17\GeV$ ($\Elab = 20\text{--}160\GeV/\text{nucleon}$). 

\subsubsection{Thermal dilepton excitation function and fireball lifetime}

A precise measurement of the thermal dilepton excitation function---the total yield of low mass thermal dileptons---is also sensitive to the fireball lifetime. It was shown in~\cite{RAPP2016586} that the integrated thermal excess radiation in $0.3<M<0.7\GeV$ tracks the total fireball lifetime remarkably well, within less than 10\%.  The NA60 measurement in \InIn at $\sqrtsNN = 17.3\GeV$ allowed the fireball lifetime to be constrained with unprecedented precision: $\tau_{\rm fb} = 7\pm1~\text{fm}/c$~\cite{RAPP2016586}. Thus, low-mass dileptons are an excellent tool to detect  {\it anomalous} variations in the fireball lifetime due, for instance, to the presence of a soft mixed phase: an increase in $\tau_{\rm fb}$ with identical final-state hadron transverse momentum spectra (\ie\ in terms of radial-flow) would necessarily imply a lifetime extension without extra collective flow, \ie\ a soft phase. 

\subsection{Quarkonium}
\label{subsec:obsquarkonium}

A suppression of heavy quarkonium states due to a colour screening mechanism has been considered, from the very beginning of this field, as one of the key signatures for the formation of a deconfined state~\cite{Matsui:1986dk}. Detailed studies, in particular for the \jpsi meson, were first performed at top SPS energy (centre of mass energy per nucleon--nucleon collision $\sqrtsNN = 17.3\GeV$)~\cite{Alessandro:2004ap}. More recently, extensive sets of measurements were obtained, for both charmonium and bottomonium states, at the RHIC ($\sqrtsNN$ up to 200\GeV) and LHC ($\sqrtsNN$ up to 5.02\TeV) ion colliders~\cite{Adare:2011yf,Adam:2016rdg}. Concerning \jpsi, a ${\sim}30$\% suppression of the \jpsi production that could not be ascribed to cold nuclear matter effects alone was observed in central \PbPb collisions at the SPS by the NA50/NA60 experiments. The size of such an ``anomalous'' suppression is qualitatively consistent with the melting in the QGP of the \chic and \psiP charmonium states, which would lead to a suppression of the \jpsi coming from the decay of those particles.

As of today, heavy quarkonium production in nucleus--nucleus collisions has not been studied below top SPS energy. NA60+ proposes to carry out a measurement of \jpsi production down to a decreased energy of the incident heavy-ion beam of approximately $\Elab = 40\GeV/\text{nucleon}$ ($\sqrtsNN =8.8\GeV$). The possible effects of the formation of a QGP with increasing \muB on charmonium states have not been thoroughly investigated by theory. On the one hand, the initial energy density of the system should decrease when moving to lower collision energies. Consequently, if the \jpsi suppression effects observed at top SPS energy are due to the dissociation of the \chic and \psiP states, one should be able to detect the beam energy threshold for the onset of their suppression. By correlating this information with the corresponding measurement of the temperature via thermal dimuons, one could experimentally identify the threshold temperature for the melting of those charmonium states. In this way a crucial test of the corresponding lattice QCD predictions can be carried out. On the other hand, the formation of a baryon-rich QGP may have unexpected, and up to now never investigated, effects even on directly produced \jpsi, making this measurement even more appealing.

By moving to lower collision energies, there are also strong indications for an increase of the size of cold nuclear matter effects on the produced \ccbar pair. Indeed, by studying \jpsi production in proton--nucleus collisions at 158 and 400\GeV incident energy~\cite{Arnaldi:2010ky}, the NA60 experiment discovered a much stronger break-up effect on the final state \jpsi at the lower of the two energies. NA60+ proposes to extend such measurements to even lower energies, with a twofold interest. First, break-up effects in cold nuclear matter are not related to QGP formation, therefore they must be corrected for when evaluating any ``anomalous'' suppression in nucleus--nucleus collisions. Therefore, such data would be mandatory for a correct interpretation of nucleus--nucleus results. Second, the production and propagation of a bound \ccbar state in cold nuclear matter is expected to be sensitive to various QCD-related phenomena, which include both initial state (nuclear shadowing) and final state effects (propagation of a colour singlet/octet pre-resonant state or the final resonance in the nucleus). Past fixed target experiments at various facilities (SPS, Tevatron, HERA) collected several sets of data, but the observations still lack a comprehensive interpretation, and new data at lower collision energy would access up to now unexplored specific kinematic configurations.

\subsection{Open charm}
\label{sec:HFmotiv}

Heavy-flavour measurements are providing unprecedented insights into the physics of hot QCD matter at small baryo\-chemical potential, as produced at the heavy-ion colliders RHIC and LHC. With a focus on the low and intermediate momentum region, measurements of \pt distributions, compared to pp collisions, and of azimuthal anisotropies of D mesons are used for example to extract fundamental transport coefficients of the QGP, such as the heavy-quark diffusion coefficient (see \eg\ Ref.~\cite{Rapp:2018qla}). Measurements of the strange-charm \Ds meson and of the \lambdac baryon, in comparison with \Dzero and \Dplus mesons containing only charm and light quarks, are used to characterize the hadronization mechanisms of charm quarks and the role of quark recombination~\cite{Plumari:2017ntm}.

It is of high interest to extend these studies into the region of finite chemical potential through heavy-ion collisions at lower energies. The charm diffusion coefficient is predicted to be larger in the hadronic phase at temperatures approaching the critical temperature $T_{\rm c}\approx 150\MeV$ from below than in the QGP phase at temperatures larger than $T_c$~\cite{He:2011yi,Scardina:2017ipo}. It should then be possible to investigate this feature in \PbPb collisions at SPS energies, where the hadronic phase $T<T_c$ represents a large part of the space-time evolution of the collisions. Measurement of \pt distributions and azimuthal anisotropy should exhibit features induced by a strong collective behaviour. In addition, such measurements that are sensitive to interactions of charm in the hadronic phase would be important input also for precision estimates of diffusion coefficients at collider energies, where a rather extensive hadronic evolution from $T_{\rm c}$ down to the kinetic freeze-out temperature of $T\sim 100\MeV$ occurs. For what concerns hadronization mechanisms, recombination effects could lead to a large enhancement of the $\rm \Lambda_c/D$ ratio. The enhancement could be larger at SPS than at RHIC and LHC energies, because of the larger baryon content of the fireball (the baryon number of the colliding nuclei is ``stopped'' in the collision region).

The total production cross section of \ccbar pairs in hadronic collisions at centre-of-mass energies below 20\GeV has never been measured with high precision, because the yields at these energies are very small. The only measurements in nucleus--nucleus collisions at the SPS were obtained by the NA60 experiment in \InIn collisions (using intermediate mass dimuons, with an uncertainty of about 20\%)~\cite{Arnaldi:2008er} and by the NA49 experiment in \PbPb collisions (upper limit using reconstructed $\rm D^0$ decays)~\cite{Alt:2005zu}. A precise measurement of this cross section in \pPb and \PbPb collisions at the SPS, besides providing the optimal normalization for \jpsi production, is also predicted to be directly sensitive to chiral symmetry restoration. The melting of the $\rm \langle \mathrm{\overline{q}q}\rangle $  condensate might also shift the threshold for $\mathrm{D}\overline{\mathrm{D}}$ production from $2\,m_{\rm D^0}\approx 3.73\GeV$ in vacuum to about 3\GeV in the chirally-symmetric medium. This reduction is predicted to lead to an enhancement of open charm meson yields by a factor of up to 7, with respect to binary scaling of the production in \pA collisions.  Thus, a large enhancement of $\mathrm{D}$ meson production is regarded as a signature for a chirally-symmetric phase~\cite{Friman:2011zz}.

Finally, measurements of D-meson production in proton--nucleus collisions at SPS energy can provide constraints to parameterisations of the nuclear modification of parton distribution functions (PDFs) at $Q^2\sim 10\text{--}40\GeV^2$ and large Bjorken-$x$ of $x_{\rm Bj} \sim 0.1\text{--}0.3$, depending on $p_{\rm T,c}\sim 0\text{--}3\GeVc$. In this kinematic region, which is poorly-constrained by existing data, the PDFs in large nuclei are expected to change from enhancement (anti-shadowing) at $x_{\rm Bj} \sim 0.1$ to suppression (``EMC effect'') at $x_{\rm Bj} \sim 0.3$ (see \eg\ Ref.~\cite{Eskola:2016oht}). NA60+ could provide precise input via ratios of the $\mathrm{D}$-meson production cross section in \pPb or \pU collisions (maximal nuclear effects) and \pBe collisions (minimal nuclear effects).

%% file: Requirements.tex
\section{Requirements (statistics, beams, installation)}
\label{sec:requirements}

The experimental programme requires to collect data at different energies (beam energy scan, BES) in the interval $\sqrtsNN \sim 5\text{--}17\GeV$.
In order to reach the required precision for the proposed measurements, the statistics goal at each energy of BES is:
\begin{itemize}[label=$\bullet$]
\item $\geq 5\cdot10^7$ reconstructed thermal muon pairs (factor 100 increase over NA60 and ${\sim} 10^5$ over RHIC/LHC experiments); 
\item $\geq 3\cdot 10^4$ reconstructed \jpsi;
\item $\geq 10^7$ reconstructed \Dzero.
\end{itemize}

The proposed physics programme of NA60+ is based on high-intensity lead beams delivered by the CERN SPS. This facility is continuously running since many years. Thanks to the new injection scheme it would be able to deliver ion beams over the required energy interval with intensities exceeding $10^7~\text{ions}/\text{s}$ (in 5 s bursts, 15 s inter-bursts). An ion beam could be delivered to a fixed target experiment while the SPS is used as ion injector for LHC. Considering the present LHC running conditions with ions, this means that ions would be available for ${\sim}4$ weeks per year.
Tentatively,  data should be collected at $\Elab = 20$, 30, 40, 80, 120 and $160\GeV/\text{nucleon}$. Further energy points might be required depending on the findings. Corresponding periods of proton beams at a few energy points will be required for reference measurements (like open charm and Drell-Yan), with intensities of ${\sim}5\cdot 10^8~\text{protons}/\text{s}$.

The  beam intensity dictates that the experiment must be installed underground. The only existing hall is the ECN3 in the CERN north area. This is presently under discussion within CERN Physics Beyond Colliders for running after LS3 during LHC Run~4 and Run~5.

%% file: Experiment.tex
\section{The NA60+ experiment, layout and detectors}
\label{sec:na60plus}

\begin{figure}[hb]
\begin{center}
\includegraphics[width=0.7\textwidth]{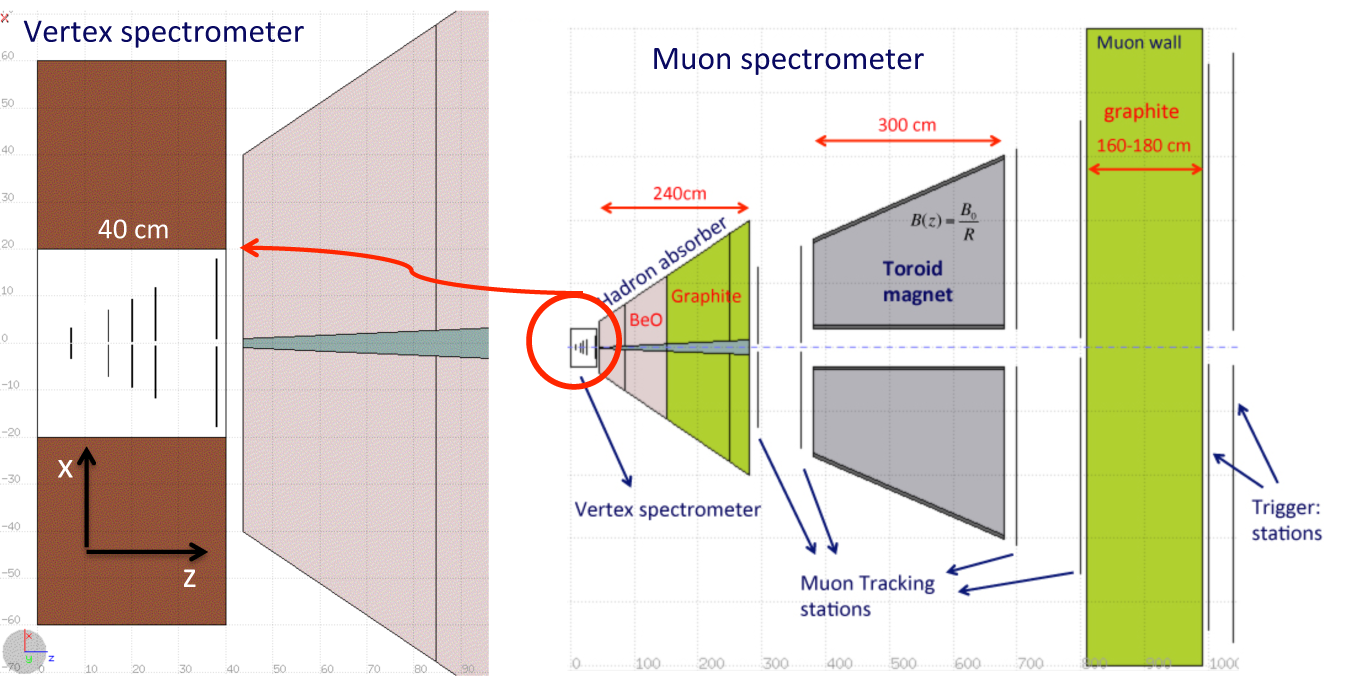}
\caption{Geometry of the proposed experimental apparatus.}
\label{fig:fig1-apparatus}
\end{center}
\end{figure}

A sketch of the NA60+ apparatus  is shown in Fig.~\ref{fig:fig1-apparatus}. Muons are measured by a magnetic spectrometer placed after a hadron absorber.  A magnet, which creates a toroidal field in the plane perpendicular to the beam direction, is placed in between the muon tracking stations and provides a field integral of 0.75~Tm at $R=1$~m. While the absorber provides the muon identification, it also degrades the kinematics of the muons, because of energy loss fluctuations and multiple scattering. This problem is overcome by measuring particle tracks also before the hadron absorber with a silicon tracker, which is the key element for a precision measurement of muons. Muon tracks are then matched to the tracks measured in the silicon vertex telescope (VT) in coordinate and momentum space. In addition, the VT provides also the measurement of the charged-particle multiplicity density, and hadronic decays of open charm.
The apparatus covers the forward rapidity hemisphere, measuring muons in the pseudo-rapidity interval $1.6<\eta<4.5$. Moreover, the apparatus is designed to guarantee a good coverage down to low transverse momenta in particular for dimuons with low mass and low pseudo-rapidity. The set-up will be adapted to the varying beam energy by scaling the absorbers thickness and moving the position of the tracking stations  in such a way to keep always a  good acceptance close to mid-rapidity.

We foresee two dedicated programmes, one with a proton beam on various nuclear targets (\pA), and one with lead beams on lead targets (\PbPb). For the \PbPb running, a segmented Pb target system of 10--15\% interaction probability (4~mm thickness) will be used, whereas for \pA collisions, different nuclear targets (Be, Al, Cu, In, W, Pb and U) with 1~mm thickness will be exposed simultaneously to the proton beam. The targets will be housed inside a vacuum box.

In the following subsections we briefly present the detector concept for the different sub-systems. Detailed performance features and simulations of the full detector response will be documented in a future Letter of Intent and Technical Design Reports at a later point.

\subsection{Detector technology}
\subsubsection{Vertex telescope: pixel detector}

The vertex telescope is a spectrometer consisting of 5 silicon pixel tracking planes immersed in a 1.2~Tm dipole field (see Fig.~\ref{fig:fig1-apparatus}). The tracker will be based on CMOS Monolithic Active Pixel Sensors (MAPS), which are now reaching impressive performances, almost ideal in many respects. At present, the state of the art is the pixel sensor ALPIDE developed for the new ALICE inner tracker. Based on the $0.18~\mu$m CMOS imaging commercial technology by TowerJazz, this $15\times30~\text{mm}^2$ sensor features a pixel size of $27\times29~\mu\text{m}^2$ with a thickness of just $50~\mu\text{m}$~\cite{AGLIERIRINELLA2017583}.
At ${\sim}1$~MHz \PbPb interaction rate, the hit flux reaches ${\sim50}$~MHz/cm$^2$ in the tracking station closest to the interaction point. The ALPIDE sensor can be operated up to ${\sim}100$~MHz/cm$^2$ but the readout rate is limited to ${\sim}10$~MHz/cm$^2$.

For NA60+ we will pursue, in close synergy with the ALICE experiment, a new generation of MAPS with almost ideal characteristics: very large area up to ${\sim}14\times14~\text{cm}^2$, retaining  thickness of 50~$\mu$m or less. The key idea is to use stitching, a technology that allows an image sensor that is larger than the field of view of the lithographic equipment to be fabricated. In this way, sensors of arbitrary size can be manufactured, the only limit being the size of the wafers.
The basic unit might be a sensor of size $1.5\times14~\text{cm}^2$~\cite{ALICE-PUBLIC-2018-013}. A resulting wafer-scale sensor could be obtained at this stage replicating this sensor chip several times along the periphery side. Such a matrix amounts to $5000\times5000$~pixels. The sensor is meant to be operated in continuous mode, recording frames at ${\sim}5\text{--}10$~MHz rate. The amount of memory buffers and serial data transmitters, all placed in the periphery, will be determined to match the required readout band-width.

All tracking stations will be identical, consisting of just few large sensors  leading to a modular, more robust and cheaper layout that allows simple replacement of non working sensors. A tracking station with few large sensors where power distribution is totally managed internally will mainly eliminate the need of a mechanical and electrical support, confining the interconnection to the outside world to the sensor edges, including the cooling. With a flex circuit for power distribution the material budget for such tracking stations might drop to a breakthrough level of 0.1\%~X$_0$.

\subsubsection{Muon tracking: GEMs}

Gas Electron Multiplier (GEM) foils~\cite{sauli1997gem} are structures commonly used as proportional counters, due to their excellent performance at high particle rates. Being successfully employed in many experiments, such as COMPASS~\cite{compassGEM}, LHCb~\cite{LHCbGEM} or TOTEM~\cite{TOTEMGEM}, GEMs have become the solution for the upcoming upgrades of the CMS Muon Endcap~\cite{CMSGEM} or the ALICE TPC~\cite{aliceTpcUpgradeTDR2014}.

GEM detectors use copper-clad polyimide foil with etched holes for gas amplification. Thanks to the photo-lithography methods used in the production process of GEM foils, the size of the amplification region could be reduced to a sub-millimetre level (typical hole diameter of 50--70~$\mu$m with a pitch of 140~$\mu$m). This allows GEMs to be operated at high particle rates of 1~MHz/cm$^{2}$~\cite{LHCbGEM} with a very good spatial resolution of $\mathcal{O}(100~\mu\text{m})$ or smaller~\cite{compassGEM}. A cascade of GEM foils permits one to attain very high proportional gains without the occurrence of discharges. Since the development and perfection of the single-mask technique~\cite{singlemask}, high-quality large-area GEMs can be produced allowing to cover substantial detector surfaces at relatively low costs. All that makes the GEM technology well suited for a high-rate muon tracker to be employed in the NA60+ spectrometer.
Following the schematic layout presented at the beginning of this section the total active area of the muon tracker is ${\sim}116$~m$^2$, organized in 4 tracking stations, behind the hadron absorber. We foresee two modules per station to improve tracking capabilities of the system (position resolution, background rejection). A single tracking module (chamber) will employ a triple GEM amplification structure with a short drift gap of $\mathcal{O}(3~\text{mm})$ and strip readout. Due to the raw-material related constraints on the maximum width of GEM foils (60~cm), we foresee a single chamber to be a ${\sim}50\times100$~cm$^2$ rectangle. Taking into account the active area of the tracker and the dimensions of a single chamber, a total number of ${\sim}$1500~GEMs will have to be produced (including spares) in order to assemble more than 450~tracking modules.
We propose to use cost-effective, non-flammable Ar-based gas mixtures with carbon dioxide (CO$_2$) as a quencher. In order to improve time resolution of the tracking module an addition of Tetrafluoromethane (CF$_4$) is considered. Both Ar/CO$_2$ and Ar/CO$_2$/CF$_4$ mixtures have been successfully used in currently operated high-rate trackers (\eg\ COMPASS or LHCb) without notable ageing effects.

Because of the large total area of the muon tracker, a pad readout would lead to an excessively large number of front-end electronics (FEE) channels. The latter can be reduced by employing 2D strip readout~\cite{compassGEM}. Given the chamber dimensions and assuming $200~\mu$m spatial resolution, a strip pitch of $\mathcal{O}(1~\text{mm})$ results in almost 900~thousand of FEE channels. Further improvement of the spatial resolution may be realized by employing a zigzag strip geometry~\cite{zigzag} or utilizing pad readout in the innermost part of the tracker.

For the readout, we consider using a VFAT chip~\cite{VFAT}, successfully employed in TOTEM and ATLAS forward detectors and planned for the CMS GEM chambers, or the new VMM3 chip~\cite{VMM3} developed for the ATLAS New Small Wheel upgrade. Detailed performance studies will be performed for the final assessment.

\subsubsection{Muon triggering: RPCs}

Resistive Plate Chambers (RPC) were used in LHC experiments for large surface tracking devices and may represent a viable choice for the muon triggering system of the NA60+ experiment. In particular, these detectors were adopted for the ALICE forward muon arm, which had a conceptual structure similar to the one proposed for NA60+. Following that set-up, one can envisage a system composed of two stations, each one featuring two RPC planes. This set-up would allow a trigger condition where one requires three out of four planes fired, providing the necessary redundancy in case of local inefficiencies in the detection planes. The single-gap detectors, with Bakelite electrodes, can be read on both sides with orthogonal strips with a pitch varying from 1 to 4 cm, following the expected occupancy and the needed spatial resolution. The detectors could be operated with a FEE which includes an amplification stage, in order to reduce the charge delivered inside the gas gap, improving both rate capability ($>100$~Hz/cm$^2$) and detector lifetime. An R\&D programme is ongoing in order to replace the flammable gas mixture used, \eg\ in ALICE with non-flammable and low-GWP gas.

A possible issue that needs investigation is related to the area of the detection planes, since, presently, the maximum size of Bakelite electrodes available is about $300\times180$~cm$^2$. With a transverse size of the detector of ${\sim}8$~m, the simple arrangement adopted in ALICE (two detectors, symmetrically positioned with respect to the beam axis) will not be possible, and some integration studies will be needed in order to devise a practical solution which ensures sufficient overlaps and avoid dead zones.
 
\subsubsection{Muon magnet}

A magnet similar to the ACM~\cite{ANDERSON198426} used by NA60 is needed with larger dimensions: length of $\sim3$ m, radius at entrance $0.3<R<1.65$~m, radius at exit $R=2.95$~m, field $B\cdot R\sim0.2\text{--}0.25~\mathrm{Tm}$. The ACM was an open toroid with field circling the beam-axis. The windings of the ACM were arranged in a complicated way, providing an excellent field quality but making this magnet difficult to build and expensive. We propose a ``minimal'' design, easier to build and cheaper, but with a more complex field~\cite{BERGSMA}. This layout could also be useful for other experiments. The main part consist of a central cylinder of 0.6~m diameter and 3~m length in the beam direction, build up out of 8~sectors. The sectors are tangentially displaced with respect to the cylinder axis in order to cross the current from one segment to another. The dimensions of the radial conductors are chosen to give little obstruction for incoming particles. The outer connection of the radial conductors is not critical. A segment consists out of a single winding, the straight conductors joined by screws. The idea is to get an easy scalable, meccano-like structure. This simple concept of minimal amount of windings has to be tested on field homogeneity and acceptance. For reasons of simplicity and cost efficiency it is proposed to add segments if the field homogeneity needs to be increased, instead of adding more outer conductors to the radial ones. Since the total current remains the same, this will not affect the acceptance. In order to reduce energy consumption the magnet will be pulsed. To avoid overheating of the central conductor, coolant has to be flown through it. It can be made laminated or foreseen of tubes. If the field quality is not satisfying due to current inhomogeneity, one can try a laminated central conductor with loop for each lamination.

%% file: PhysicsPerformances.tex
\section{Physics performances}
\label{sec:physics_performances}

\subsection{Thermal dileptons: caloric curve, \texorpdfstring{$\rho\text{--}\aone$}{rho--a1} chiral mixing, excitation function}
\label{subsec:ThermalRadiationPerformance}

Detailed performance studies  were carried out for 5\% most central \PbPb collisions at $\sqrtsNN = 6.3$, 8.8 and 17.3~GeV. The differential spectra of thermal \mumu pairs, $\dd^3 N/(\dd M \dd \pt \dd y)$, are based on the in-medium $\rho$, $\omega$ and  $4\pi$ spectral functions, QGP and the expanding thermal fireball model of~\cite{RAPP2016586}. The generator is based on the model calculation which assumes either no $\rho\text{--}\aone$ chiral mixing or full chiral mixing ($\epsilon=1/2$) in the mass region $1<M<1.5\GeV$. For the performance of the temperature measurement, thermal dileptons were generated without chiral mixing.
The hadron cocktail generator for the 2-body decays of $\eta$, $\omega$,  and $\phi$ and the Dalitz decays  $\eta\to\gamma\mu^+\mu^-$ and $\omega\to\pi^0\mu^+\mu^-$ is based on the NA60 generator and on the statistical model of~\cite{Becattini:2005xt}. The Drell-Yan process and open-charm production are simulated with the \pythia event generator.

\begin{figure}[ht]
\begin{center}
\includegraphics[width=0.9\textwidth]{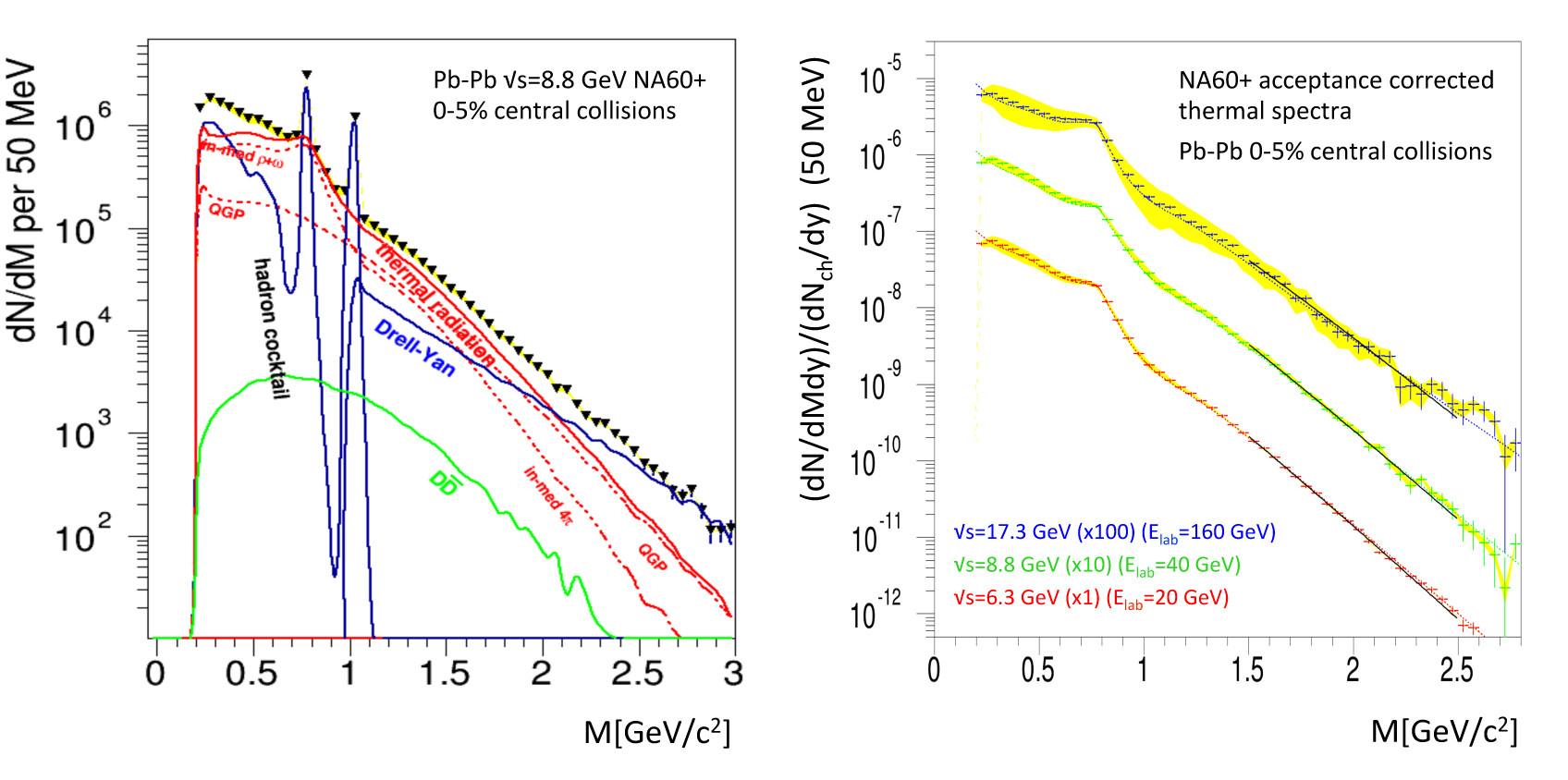}
\caption{(Left) expected signal sample in 5\% most central \PbPb collision at $\sqrtsNN = 8.8\GeV$ after subtraction of combinatorial and fake match background. (Right) acceptance corrected thermal spectra at three different beam energies obtained after subtraction of open charm, Drell-Yan and hadronic cocktail below 1\GeV.}
\label{fig:fig1-thermal-performance}
\end{center}
\end{figure}

We present the results for a sample of $2\cdot10^7$ reconstructed pairs in central collisions at each energy, corresponding to a total sample of ${\sim}5\cdot10^7$ integrated in centrality at each energy. The left panel of Fig.~\ref{fig:fig1-thermal-performance} shows the signal reconstructed mass spectra (black) for $\sqrtsNN = 8.8\GeV$ after subtraction of the combinatorial background due to pion and kaon decays as well as fake matches. The latter arise from incorrect matches of a muon track to a track reconstructed in the VT. The combinatorial background is subtracted assuming a 0.5\% systematic uncertainty (shown as a yellow band). For what concerns minimum bias collisions, the progress in statistics over NA60 is a factor ${\sim}100$, retaining a similar background ratio and a better mass resolution. At a ${\sim}1$~MHz interaction rate, this statistics can be collected in a ${\sim}30$~days run.
The figure shows all the expected signal components. For $M<1\GeV$, the thermal radiation yield is dominated by the in-medium $\rho$. The $\omega$ and $\phi$ peaks are well resolved with a resolution of ${\sim}10\MeV$ at the $\omega$ mass.  The  thermal spectrum is measurable up to 2.5--3\GeV. The open charm yield becomes totally negligible at low energy. The Drell-Yan yield will be measured in dedicated \pA runs (see also section~\ref{subsec:JpsiPerformance}).
The thermal spectra are obtained after (i) subtraction of the hadronic cocktail for $M<1\GeV$ of $\phi$, $\omega$ and $\eta$ decays into \mumu and the $\omega$, $\eta$ Dalitz decays and (ii) subtraction of Drell-Yan and open charm muon pairs for $M>1\GeV$. After acceptance correction the spectra are fit with $\dd N/\dd M\propto M^{3/2}\exp(-M/T_s)$ in the interval $M=1.5\text{--}2.5\GeV$. This is shown in the right panel of Fig.~\ref{fig:fig1-thermal-performance}.

\begin{figure}[ht]
\begin{center}
\includegraphics[width=1.0\textwidth]{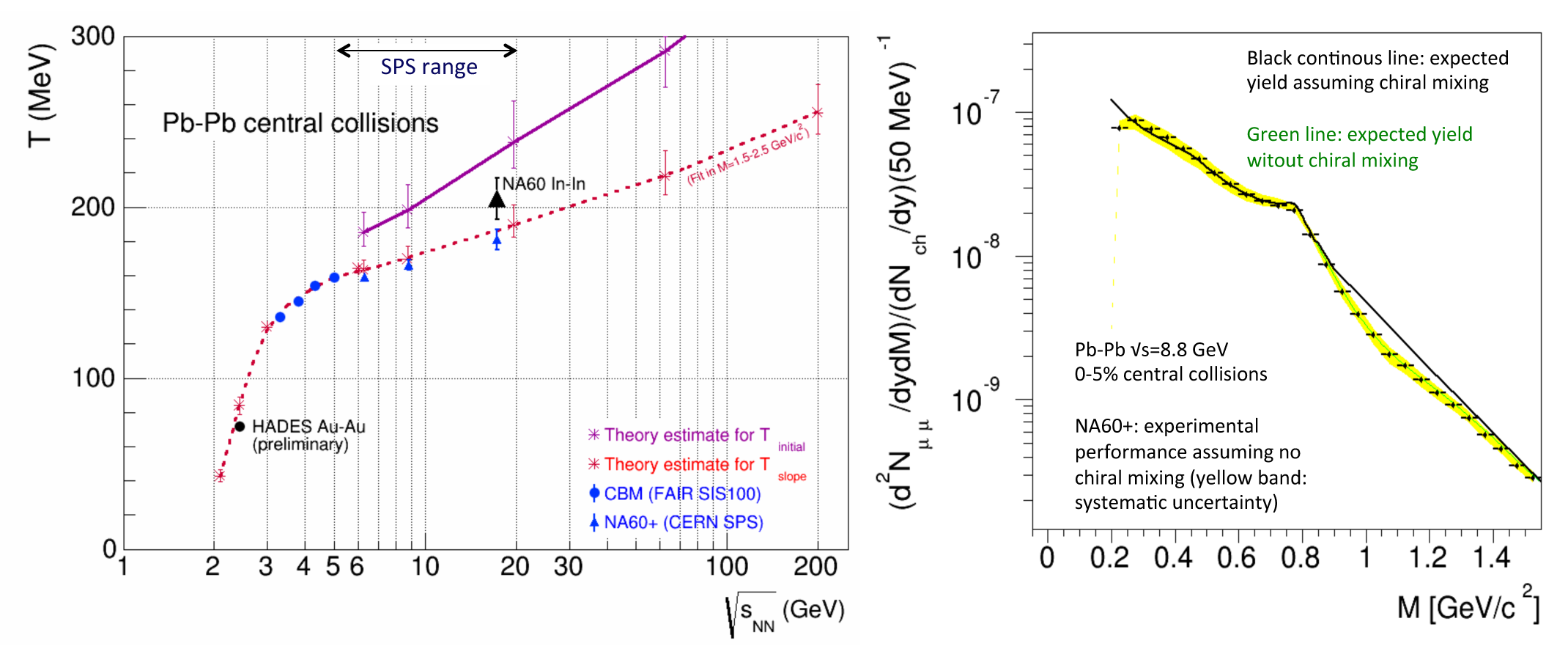}
\caption{(Left) medium temperature evolution vs $\sqrtsNN$ in central \PbPb collisions. $T_{\rm initial}$ (magenta) and $T_{\rm slope}$ (red) are theoretical estimates for the initial medium temperature and the temperature from dilepton spectra respectively, using~\cite{RAPP2016586} and a coarse graining approach in UrQMD~\cite{PhysRevC.92.014911,GALATYUKQM2018}. Blue triangles  are the expected performance from NA60+ (CBM performance is also shown). The only existing measurements at present are from NA60 \InIn~\cite{Arnaldi:2008er,Specht:2010xu} and from HADES \AuAu~\cite{HADESQM2018}. (Right) NA60+ projection for the acceptance corrected thermal dimuon mass spectrum at $\sqrtsNN = 8.8\GeV$ in case of no chiral mixing compared to the theoretical expectation (green line). The black line above 1\GeV is the expectation from full chiral mixing~\cite{RAPP2016586}.}
\label{fig:fig2-thermal-performance}
\end{center}
\end{figure}

The resulting temperatures are compared to the theoretical model used as an input in the left panel of Fig.~\ref{fig:fig2-thermal-performance}. At low energies, the temperatures can be measured with a combined statistical and systematic uncertainty of just a few\MeV, thus showing that the experiment has a strong sensitivity to a possible flattening of the caloric curve.
The acceptance corrected mass spectrum at $\sqrtsNN = 8.8\GeV$, based on the assumption of no chiral mixing, is compared to the expectation in case of full chiral mixing in the right panel of Fig.~\ref{fig:fig2-thermal-performance}. As shown, the statistical and systematic uncertainty provide a very good sensitivity to an increase of the yield due to chiral mixing of ${\sim}20\text{--}30\%$.


\begin{figure}[ht]
\begin{center}
\includegraphics[width=0.45\textwidth]{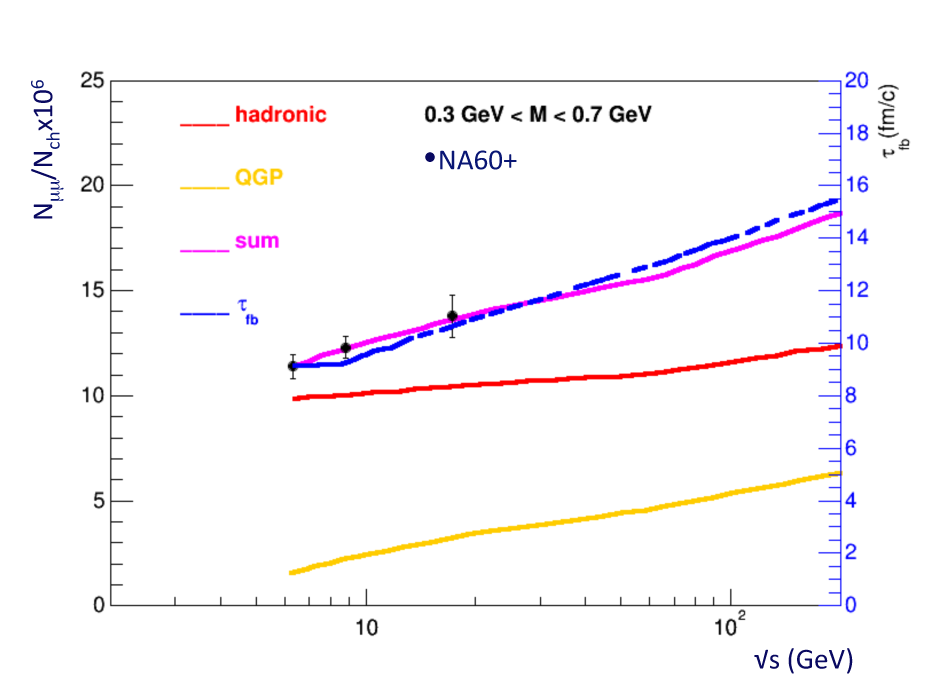}
\caption{Performance for thermal dilepton yield measurement for the fireball lifetime measurement.}
\label{fig:fig3-thermal-performance}
\end{center}
\end{figure}

Finally the performance for the dilepton excitation function, \ie\ the total thermal yield measurement in the mass range 0.3--0.7\GeV, is compared to the lifetime estimate of~\cite{RAPP2016586} in Fig.~\ref{fig:fig3-thermal-performance}. The  uncertainty is dominated by the systematic error from background subtraction. The measurement at low energies provides very good sensitivity to possible anomalies in the fireball lifetime.

\subsection{Open charm measurements: hadronic decays of mesons/baryons}

Open charm hadrons can be fully reconstructed with the NA60+ apparatus via their decays into two or three charged hadrons. In particular, the following decays could enable the measurement of non-strange and strange D mesons as well as $\lambdac$ baryons: $\Dzero \to {\rm K}^-\pi^+$, $\Dplus\to {\rm K}^-\pi^+\pi^+$, ${\Ds \to \phi\pi^+\to {\rm K}^{+} {\rm K}^{-} \pi^{+}}$, $\lambdacplus \to {\rm p K^{-}}\pi^+$ and their charge conjugates. The decay particles (pions, kaons and protons) are detected by reconstructing their tracks in the silicon-pixel detectors of the VT. D-meson and \lambdac candidates are built combining pairs or triplets of tracks with the proper charge signs. The huge combinatorial background can be reduced via geometrical selections on the displaced decay-vertex topology, exploiting the fact that the mean proper decay lengths $c\tau$ of open charm hadrons are of about 60--310~$\mu$m, and therefore their decay vertices are typically displaced by a few hundred~$\mu$m from the interaction vertex. Among the measurements proposed by NA60+, open-charm hadron studies are those that impose the strongest constraints on the design of the VT detectors, which should provide good resolution on the track parameters in order to allow us to separate the secondary tracks produced in open-charm hadron decays from the primary ones originating from the interaction point.

The benchmark studies were carried out for the measurement of $\Dzero \to {\rm K}^-\pi^+$ in the 5\% most central \PbPb collisions at two different beam energies: 158 and $60\GeV/\text{nucleon}$, corresponding to $\sqrtsNN = 17.3$ and 10.6\GeV, respectively. The \Dzero and $\overline{\Dzero}$ mesons were simulated with \pt and rapidity distributions obtained with the {\sc Powheg-Box} event generator~\cite{Alioli:2010xd} for the hard-scattering and \pythia~6 for the parton shower and hadronization. The combinatorial background was estimated by simulating pions, kaons and protons with multiplicity, \pt and rapidity distributions taken from the parameterisations published by the NA49 collaboration in Refs.~\cite{Afanasiev:2002mx,Alt:2006dk}. The number of particles per event at $\sqrtsNN = 17.3\GeV$ is about 1200, which produce about 350k background candidates per event, out of which about 8000 have an invariant mass within 60\MeV from the \Dzero-meson mass (1.865\GeV). The yield per event at $\sqrtsNN = 17.3\GeV$, estimated assuming  $\sigma_{\ccbar}=5~\mu$b (based on~\cite{Lourenco:2006vw,Vogt:2001nh} and \powheg) and $f({\rm c}\to \Dzero)\sim0.55$, is about 0.006. The signal-to-background ratio is therefore about $7\cdot 10^{-7}$ and needs to be enhanced with the kinematical and geometrical selections.

\begin{figure}[ht]
\begin{center}
\includegraphics[width=0.85\linewidth]{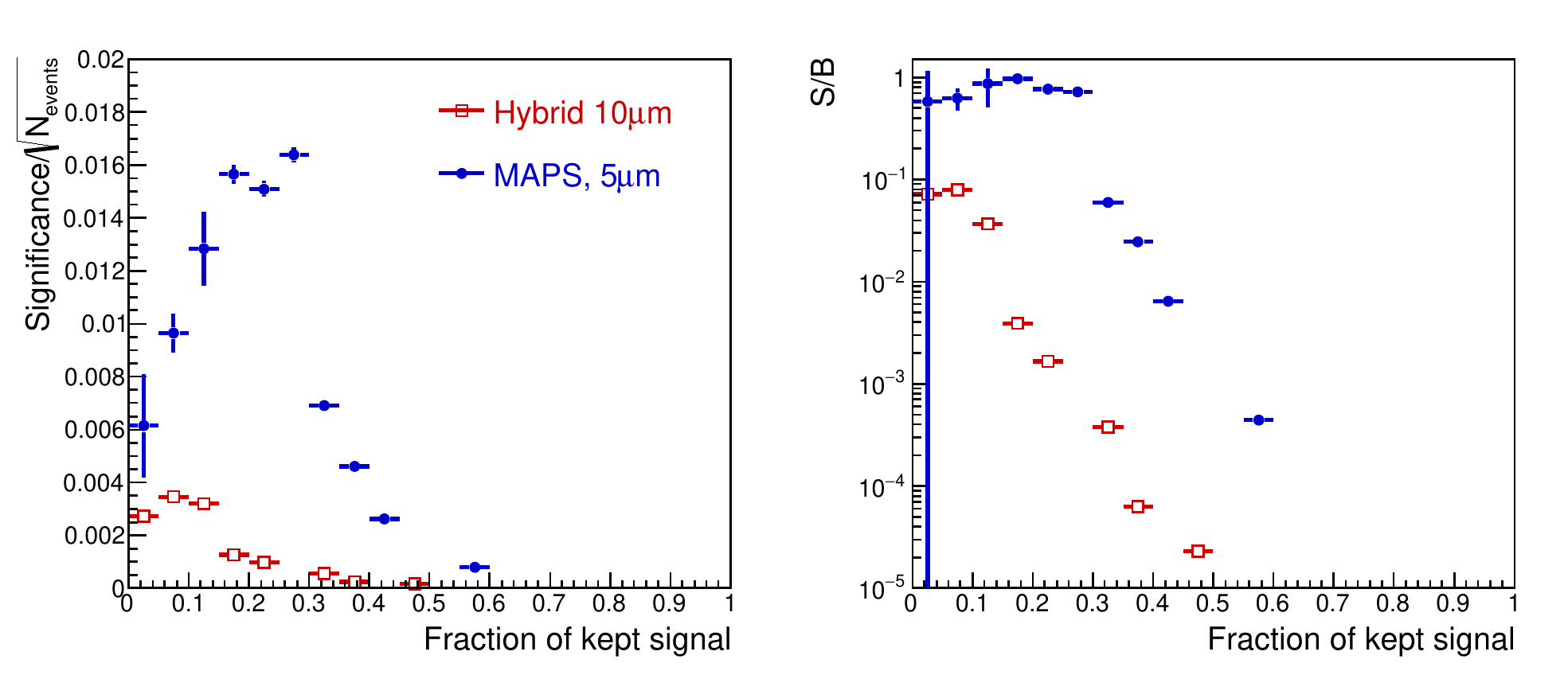}
\caption{\Dzero-meson statistical significance (normalized to the square root of the number of events) and signal-over-background ratio as a function of the selection efficiency for the 5\% most central \PbPb collisions at $\sqrtsNN = 17.3\GeV$. The signal and the background yields are evaluated in a $2\sigma$ invariant-mass region around the \Dzero peak.}
\label{fig:D0signif}
\end{center}
\end{figure}

\begin{figure}[ht]
\begin{center}
\includegraphics[width=0.45\linewidth]{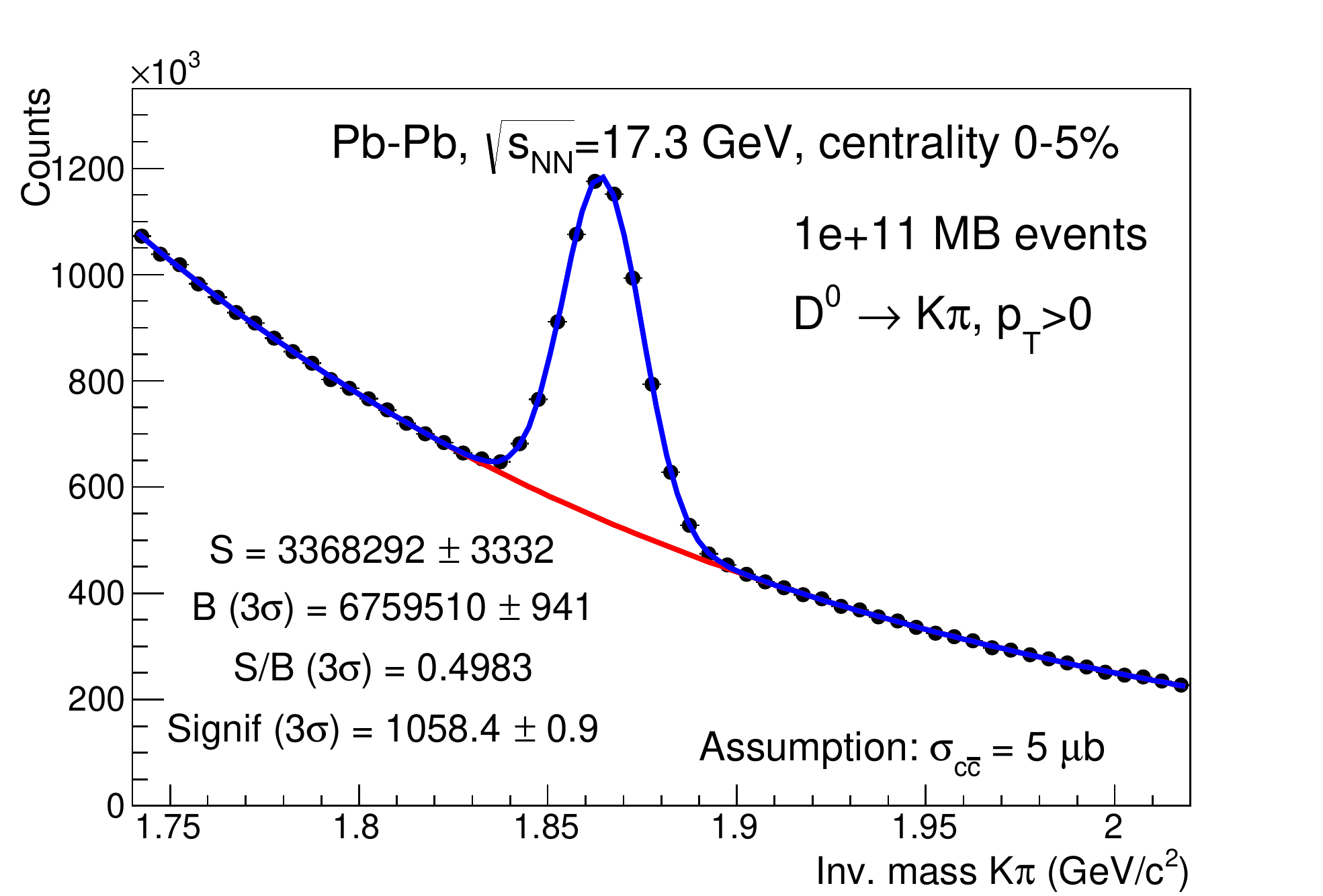}
\includegraphics[width=0.45\linewidth]{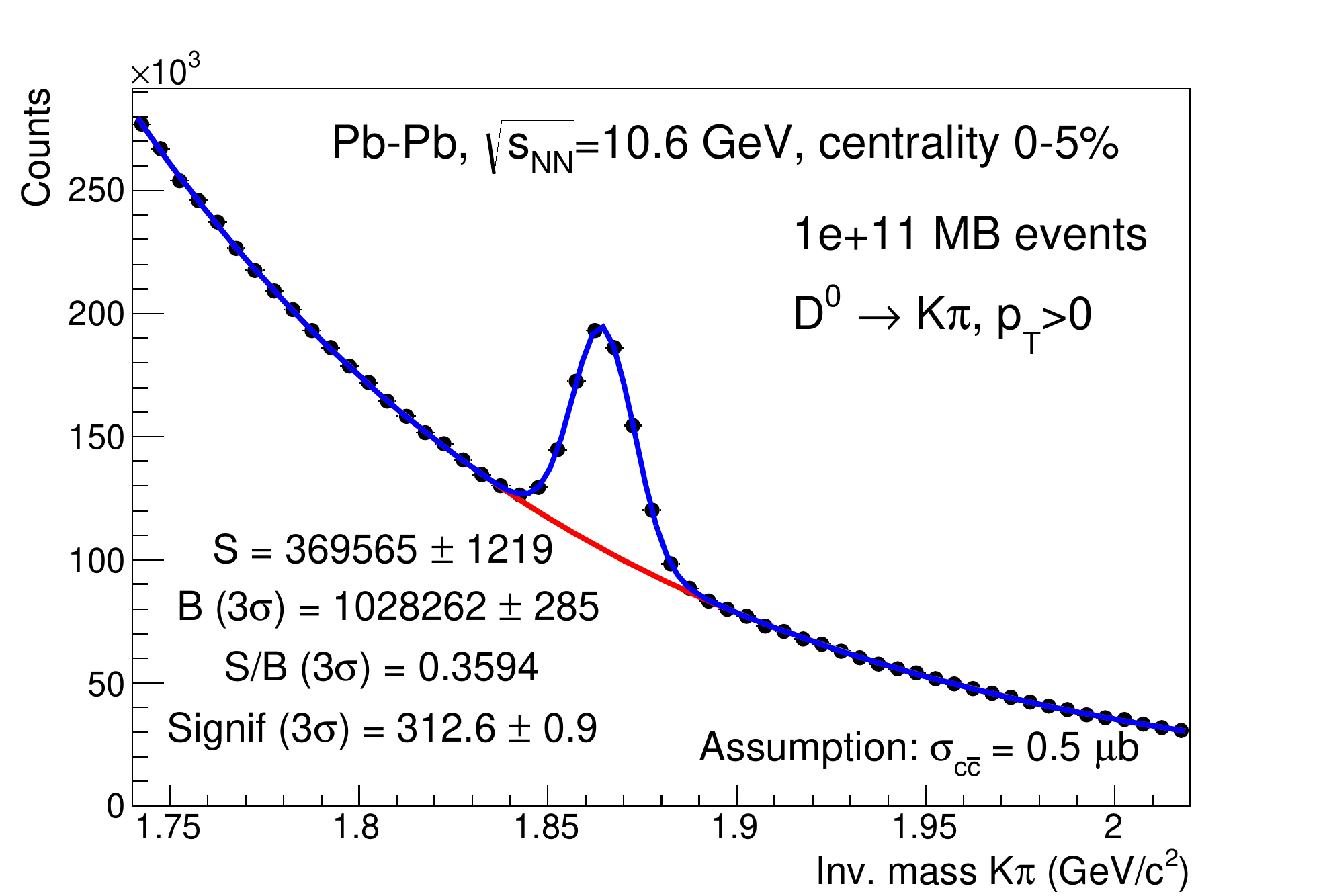}
\caption{Projection for invariant-mass distribution of \Dzero candidates in $5\cdot 10^{9}$ central \PbPb collisions at beam energies of 160 (left) and 60 (right) $\GeV/\text{nucleon}$ for the case of the VT detector based on MAPS with 5~$\mu$m spatial resolution.}
\label{fig:D0invmass}
\end{center}
\end{figure}

Several selection sets were tested, and for each of them the signal-selection efficiency, the statistical significance of the signal ($S/\sqrt{S+B}$) and the signal-to-background ratio ($S/B$) were computed. In Fig.~\ref{fig:D0signif}, the significance per event and the $S/B$ at $\sqrtsNN = 17.3\GeV$ are shown as a function of the efficiency for two different configurations of the VT detectors, namely the one based on hybrid pixel sensors with $10~\mu$m spatial resolution and the one based on Monolithic Active Pixels (MAPS) which feature a better spatial resolution ($5~\mu$m) and a reduced material budget. The performance is substantially better with the MAPS detector, which provides a better resolution on the decay track momentum, on the decay vertex position (10--15~$\mu$m in the plane transverse to the beam line with MAPS vs. 30--40~$\mu$m with hybrid pixels) and therefore on the \Dzero invariant mass (10\MeV vs. 24\MeV). In the left panel of Fig.~\ref{fig:D0invmass}, we show a projection for the invariant-mass distribution of \Dzero candidates in $5\cdot 10^{9}$ central \PbPb collisions at $\sqrtsNN = 17.3\GeV$, corresponding to a sample of $10^{11}$~minimum bias (MB) collisions which can be collected in one month of data taking at 150~kHz of acquisition rate. The MAPS detector would enable a measurement of the \Dzero-meson yield in central \PbPb collision with a statistical precision much better than 1\%, which would allow also for studies in \pt and $y$ intervals and for the determination of the elliptic flow coefficient $v_2$ of D mesons with percent level statistical uncertainty. In the right panel of Fig.~\ref{fig:D0invmass}, the projected performance for the 5\% most central \PbPb collisions at $\sqrtsNN = 10.6\GeV$ is shown for the case of MAPS detectors. For this performance study we assumed $\sigma_{\ccbar}=0.5~\mu$b and the combinatorial background was simulated based on interpolation of NA49 measurements at 40 and $80\GeV/\text{nucleon}$ incident energy. It demonstrates that the integrated \Dzero-meson cross section can be measured with statistical precision better than 1\% at collision energies at which the charm cross section is poorly known experimentally.

The excellent performance obtained for the two-body decays of \Dzero mesons suggests that open-charm measurements will also be feasible for three-body decay channels. This opens the possibility to measure the production yield of strange D mesons and of \lambdac baryons.

\subsection{\texorpdfstring{\jpsi}{J/psi} suppression: onset of deconfinement}
\label{subsec:JpsiPerformance}

\begin{figure}[ht]
\begin{center}
\includegraphics[width=0.45\linewidth]{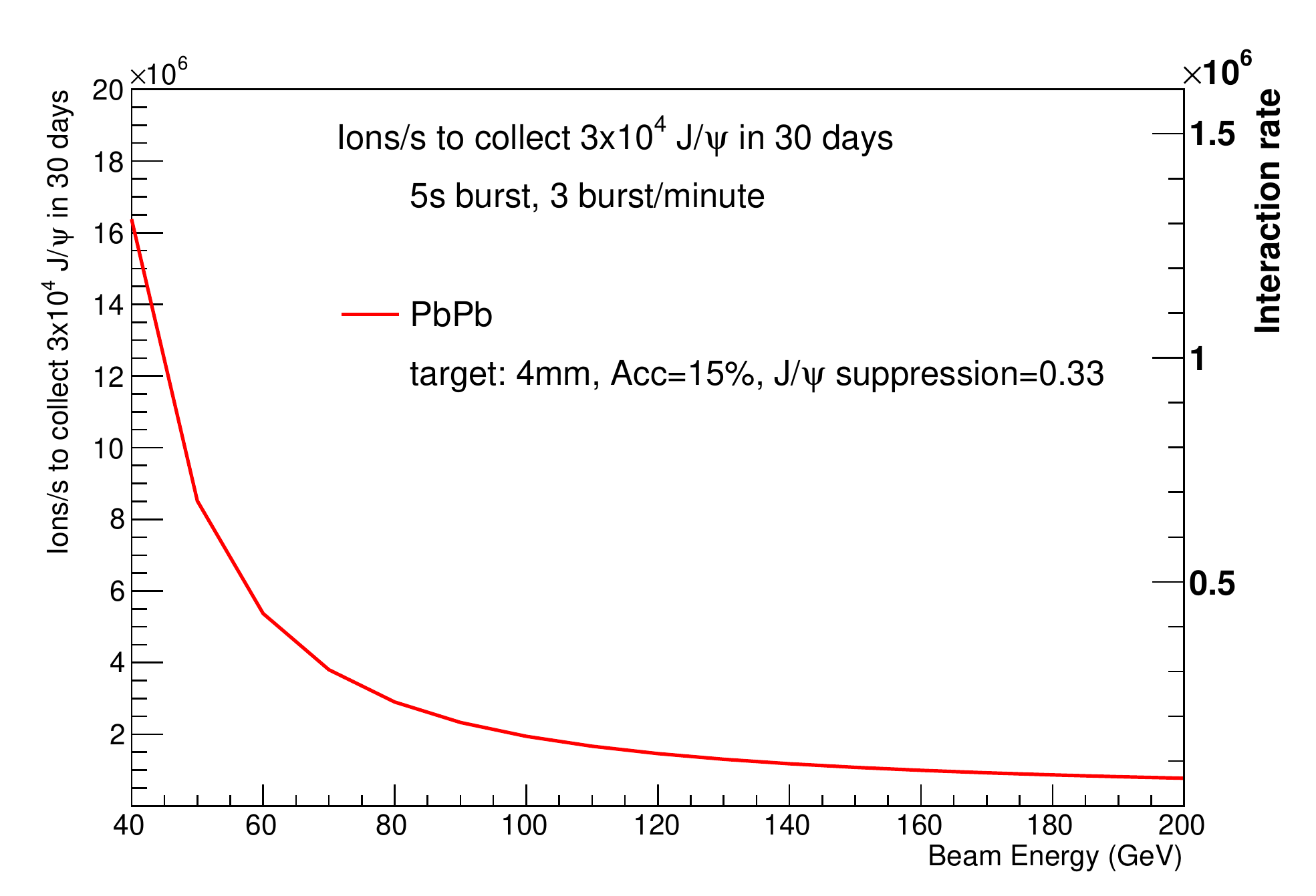}
\caption{Interaction rate and beam intensity needed to collect a sample of $3\cdot$10$^4$ \jpsi in \PbPb collisions at various energies.}
\label{fig:Jpsiperf}
\end{center}
\end{figure}

Among the measurements proposed by NA60+, quarkonium studies require the highest integrated luminosity $L_{\rm int}$, due to the small production cross section at low incident beam energy. On the other hand, past SPS experiments (NA50/NA60) showed that the background levels in the dimuon invariant mass spectrum in the \jpsi region are also very small (${<}5\%$)~\cite{Arnaldi:2007zz}. In addition, the NA60 experiment was able to perform a very accurate measurement of the centrality dependence of the suppression by collecting ${\sim}3\cdot 10^4$ \jpsi in \InIn collisions at $158\GeV/\text{nucleon}$ incident energy, \ie\ $\sqrtsNN = 17.3\GeV$. In order to assess the possibility of a measurement with similar accuracy at lower incident energy, we have calculated the interaction rate needed in order to collect $3\cdot 10^4$ \jpsi in a 30-day long \PbPb data taking period with NA60+. We have assumed (i) SPS burst structure similar to the one at top SPS energy, (ii) realistic values for the acceptance of the apparatus in the rapidity region $|y|<0.5$ and (iii) a suppression of the \jpsi leading to a $2/3$ reduction of the expected yields. Assuming a target thickness of 4~mm, we have also evaluated the corresponding beam intensity needed to collect the desired statistics. The result, shown in Fig.~\ref{fig:Jpsiperf}, shows that a large \jpsi sample can be collected while keeping the interaction rate below 1~MHz, corresponding to a beam intensity of about $10^7$ Pb ions/s.

\begin{figure}[ht]
\begin{center}
\includegraphics[width=0.4\linewidth]{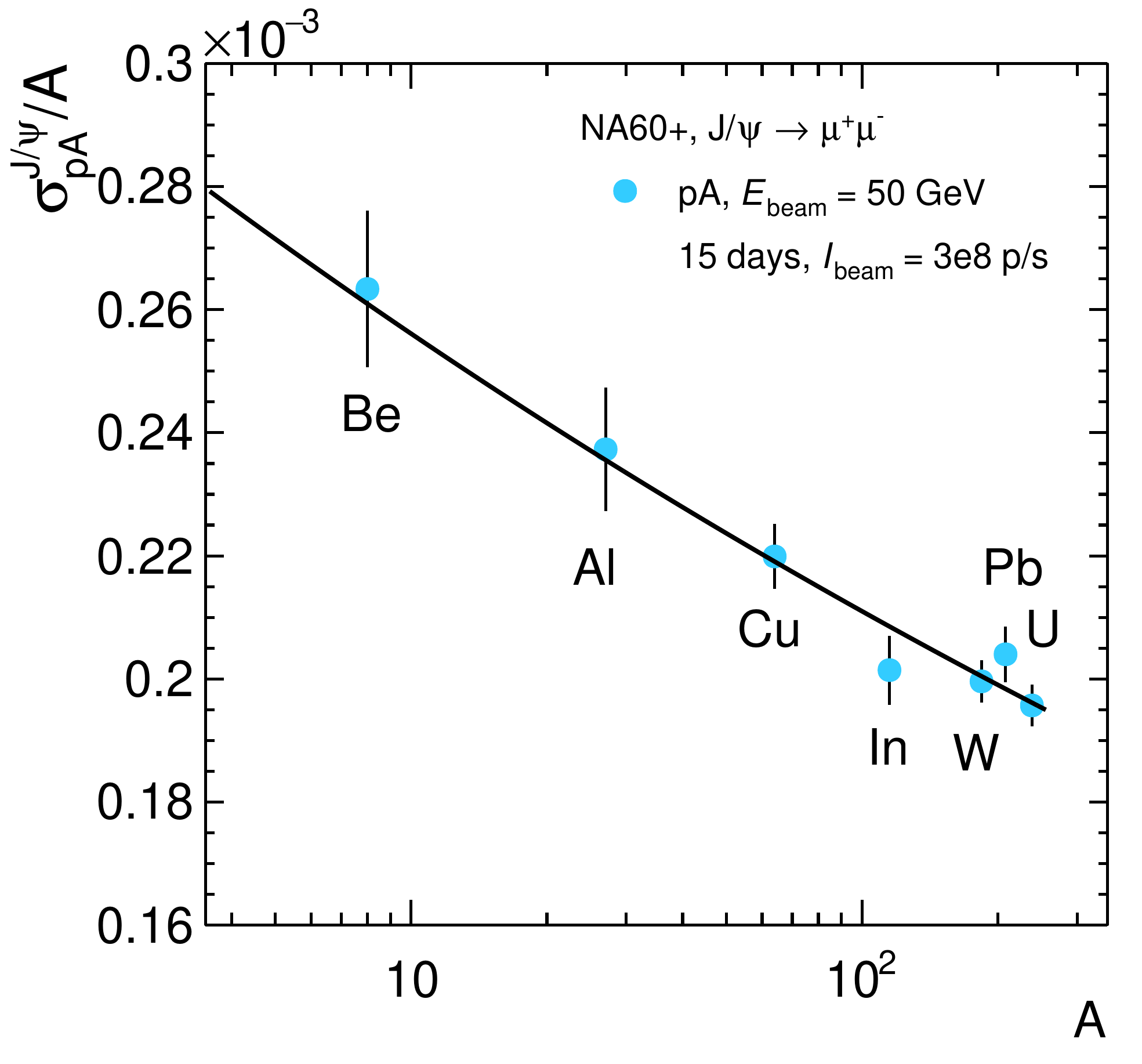}
\includegraphics[width=0.4\linewidth]{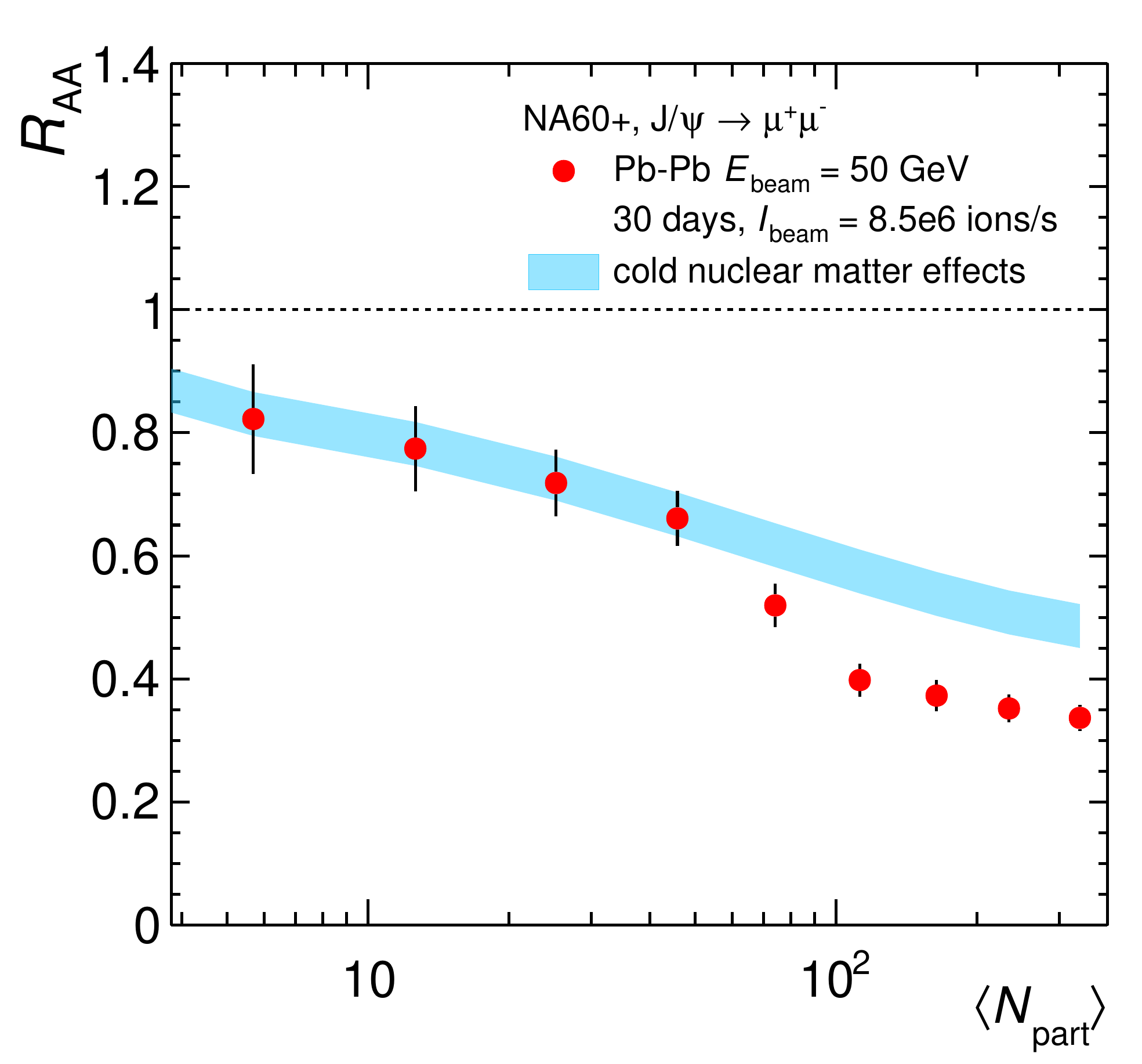}
\caption{(Left) \jpsi cross section normalized to the mass number A for \pA collisions at $\Elab = 50\GeV$. The results of a fit using the parameterisation $\sigma^{\jpsi}_{\pA}=\sigma^{\jpsi}_{\pp}\cdot A^{\alpha}$ are shown. (Right) The nuclear modification factor for \jpsi production in \PbPb collisions at $\Elab = 50\GeV$ as a function of \Npart, compared with expectations from cold nuclear matter effects, obtained from the \pA results and shown as a blue band.}
\label{fig:Jpsiresult}
\end{center}
\end{figure}

At SPS energies, it is well known that the break-up of the \jpsi in cold nuclear matter plays an important role in determining the final observed yields in nucleus--nucleus collisions. Therefore, data taking with \pA collisions are mandatory to calibrate such an effect. We assumed to have 7 nuclear targets as mentioned in section~\ref{sec:na60plus} and 15 days of data taking at a beam intensity of $3\cdot 10^8~\text{protons}/\text{s}$. In the left panel of Fig.~\ref{fig:Jpsiresult}, the expected cross sections as a function of the mass number A are shown, for an incident beam energy $\Elab = 50\GeV$, or $\sqrtsNN = 9.8\GeV$, and in the hypothesis of a dissociation cross section $\sigma_{\jpsi\text{--}\rm N}=4.3$~mb. Such results are necessary in order to (i) evaluate the \jpsi production cross section $\sigma_{\pp\rightarrow \jpsi \rm X}$, needed for \raa, (ii) extrapolate break-up effects to the geometrical conditions of \PbPb results. These procedures were well tested in past SPS experiments~\cite{Alessandro:2003pi}.

Finally, the right panel of Fig.~\ref{fig:Jpsiresult} shows the results of a simulation of the \jpsi\ \raa as a function of the number of participant nucleons, \Npart, assuming that for $\Npart \lesssim 50$ the suppression is entirely due to cold nuclear matter effects, while for more central events an extra-suppression reaching 20\% sets in. The simulation assumes 30~days of \PbPb collisions at $\Elab = 50\GeV$, with an SPS beam intensity of $8.5\cdot 10^6~\text{Pb ions}/\text{s}$. The uncertainties shown for \raa in the right panel of Fig.~\ref{fig:Jpsiresult} include, in addition to the statistical uncertainty on the \jpsi, conservatively evaluated assuming a 20\% background level, those on the evaluation of the centrality variables (Glauber model) and on the \pp cross section, calculated from the \pA results displayed in the left panel of Fig.~\ref{fig:Jpsiresult}. One can clearly see that a precise evaluation of a relatively small anomalous \jpsi suppression is within reach down to low \Elab.

Corresponding studies for the detection of the $\psiP\rightarrow\mumu$ and $\chic\rightarrow\mumu\gamma$ decays are ongoing and are expected to nicely complement the physics programme of NA60+ in the charmonium sector.

%% file: Competition.tex
\section{Competition with present and future measurements}
\label{sec:competition}

Several proposals have been put forward to investigate experimentally the QCD phase diagram at high \muB in the next decade. Since thermal dilepton radiation, open charm and quarkonia are produced by rare processes, high precision measurements require very large interaction rates and operation in fixed target mode. Moreover, the \muB interval accessible by a single experiment should be as wide as possible. The combination of these two key aspects is such that the physics programme described in this document can be uniquely pursued at the CERN SPS. Although other facilities for the study of ultra-relativistic heavy-ion collisions in the high-\muB domain are being built (FAIR, NICA) they are either lacking the necessary interaction rate (NICA) or reach too low energy (SIS100 at FAIR) to address all the physics topics addressed by NA60+. Similarly, the beam energy scan programme at RHIC (BES-II) covers the same energy range as the CERN SPS, but with interaction rates lower by orders of magnitude. It should be noted that NA60+ and CBM with their high-interaction rates but different collision energies would nicely complement each other in mapping out, \eg\ the caloric curve.

%% file: Summary.tex
\section{Conclusions}
\label{sec:summary}

The exploration of the high-\muB region of the QCD phase diagram represents one of the main directions in the evolution of the field of ultrarelativistic heavy-ion collisions. This document, submitted to the European Strategy for Particle Physics, summarizes the status of the NA60+ project, which aims at performing a comprehensive measurement, in the energy range accessible to the CERN SPS, of hard and electromagnetic processes, with unprecedented accuracy. The addressed observables include the production of thermal muon pairs from the QGP phase, the investigation of the modifications of the vector meson spectral functions as well as the study of charmonia and open charm mesons/baryons. Preliminary physics performance studies discussed in this document show that a set-up which makes use of existing advanced detection techniques would be able to carry out the proposed physics programme, provided that the experiment can profit of high-intensity ion beams as those that can be delivered in the ECN3 experimental hall.

%% file: authorlist-preprint.tex
\begingroup
\small
\begin{flushleft}
M.~Agnello\Irefnn{torino}{disat}\And
F.~Antinori\Irefn{padova}\And
H.~Appelsh\"{a}user\Irefn{frankfurt}\And
R.~Arnaldi\Irefn{torino}\And
R.~Bailhache\Irefn{frankfurt}\And
L.~Barioglio\Irefnn{utorino}{torino}\And
S.~Beole\Irefnn{utorino}{torino}\And
A.~Beraudo\Irefn{torino}
A.~Bianchi\Irefnn{utorino}{torino}\And
L.~Bianchi\Irefnn{utorino}{torino}\And
E.~Botta\Irefnn{utorino}{torino}\And
E.~Bruna\Irefn{torino}\And
S.~Bufalino\Irefnn{disat}{torino}\And
E.~Casula\Irefnn{cagliari}{ucagliari}\And
F.~Catalano\Irefnn{disat}{torino}\And
S.~Chattopadhyay\Irefn{saha}\And
A.~Chauvin\Irefn{cagliari}\And
C.~Cicalo\Irefn{cagliari}\And
M.~Concas\Irefnn{det}{torino}\And
P.~Cortese\Irefnn{piemonte}{torino}\And
T.~Dahms\Irefnn{cluster}{tum}\Aref{editor}\And
A.~Dainese\Irefn{padova}\And
A.~Das\Irefn{saha}\And
D.~Das\Irefn{saha}\And
D.~Das\Irefn{saha}\And
I.~Das\Irefn{saha}\And
L.~Das Bose\Irefn{saha}\And
A.~De Falco\Irefn{cagliari}\And
N.~De Marco\Irefn{torino}\And
S.~Delsanto\Irefnn{utorino}{torino}\And
A.~Drees\Irefn{stonybrook}\And
L.~Fabbietti\Irefn{tum}\And
P.~Fecchio\Irefnn{disat}{torino}\And
A.~Ferretti\Irefnn{utorino}{torino}\And
A.~Feliciello\Irefn{torino}\And
M.~Gagliardi\Irefnn{utorino}{torino}\And
P.~Gasik\Irefn{tum}\And
F.~Geurts\Irefn{rice}\And
P.~Giubilato\Irefnn{padova}{upadova}\And
V.~Greco\Irefn{catania}\And
F.~Grosa\Irefnn{disat}{torino}\And
H.~Hansen\Irefn{lyon}\And
J.~Klein\Irefn{torino}\And
W.~Li\Irefn{rice}\And
M.P.~Lombardo\Irefn{lfn}
M.~Masera\Irefnn{utorino}{torino}\And
A.~Masoni\Irefn{cagliari}\And
L.~Micheletti\Irefnn{utorino}{torino}\And
L.~Musa\Irefn{cern}\And 
M.~Nardi\Irefn{torino}\And
H.~Onishi\Irefn{tohoku}\And
C.~Oppedisano\Irefn{torino}\And
B.~Paul\Irefnn{cagliari}{ucagliari}\And
S.~Plumari\Irefn{ucatania}\And
F.~Prino\Irefn{torino}\And
M.~Puccio\Irefnn{utorino}{torino}\And
L.~Ramello\Irefnn{piemonte}{torino}\And
R.~Rapp\Irefn{tamu}\And
I.~Ravasenga\Irefnn{disat}{torino}\And
A.~Rossi\Irefnn{padova}{upadova}\And
P.~Roy\Irefn{saha}\And
E.~Scomparin\Irefn{torino}\Aref{editor}\And
S.~Siddhanta\Irefn{cagliari}\And
R.~Shahoyan\Irefn{cern}\And
T.~Sinha\Irefn{saha}\And
M.~Sitta\Irefnn{piemonte}{torino}\And
H.~Specht\Irefn{heidelberg}\And
S.~Trogolo\Irefnn{utorino}{torino}\And
R.~Turrisi\Irefn{padova}\And
A.~Uras\Irefn{lyon}\And
G.~Usai\Irefnn{cagliari}{ucagliari}\Aref{editor}\Aref{contact}\And
E.~Vercellin\Irefnn{utorino}{torino}\And
J.~Wiechula\Irefn{frankfurt}
\renewcommand\labelenumi{\textsuperscript{\theenumi}~}

\subsection*{Notes}
\renewcommand\theenumi{\roman{enumi}}
\bigskip
\begin{Authlist}
\item \Adef{editor}Editors
\item \Adef{contact}E-mail: \href{mailto:gianluca.usai@ca.infn.it}{gianluca.usai@ca.infn.it}
\end{Authlist}

\bigskip
\subsection*{Collaboration Institutes}
\renewcommand\theenumi{\arabic{enumi}~}
\bigskip
\begin{Authlist}
\item \Idef{lyon}Universit\'{e} de Lyon, Universit\'{e} Lyon 1, CNRS/IN2P3, IPN-Lyon, Villeurbanne, Lyon, France
\item \Idef{frankfurt}Institut f\"{u}r Kernphysik, Johann Wolfgang Goethe-Universit\"{a}t Frankfurt, Frankfurt, Germany
\item \Idef{heidelberg}Physikalisches Institut, Ruprecht-Karls-Universit\"{a}t Heidelberg, Heidelberg, Germany
\item \Idef{cluster}Excellence Cluster `Universe', Technische Universit\"{a}t M\"{u}nchen, Munich, Germany
\item \Idef{tum}Physik Department, Technische Universit\"{a}t M\"{u}nchen, Munich, Germany
\item \Idef{saha}Saha Institute of Nuclear Physics, Homi Bhabha National Institute, Kolkata, India
\item \Idef{cagliari}INFN, Sezione di Cagliari, Cagliari, Italy
\item \Idef{ucagliari}Dipartimento di Fisica dell'Universit\`{a} di Cagliari, Cagliari, Italy
\item \Idef{catania}INFN, Laboratori Nazionali del Sud, Catania, Italy
\item \Idef{ucatania}Dipartimento di Fisica e Astronomia dell'Universit\`{a} di Catania, Catania, Italy
\item \Idef{lfn}INFN, Laboratori Nazionali di Frascati, Frascati, Italy
\item \Idef{padova}INFN, Sezione di Padova, Padova, Italy
\item \Idef{upadova}Dipartimento di Fisica e Astronomia dell'Universit\`{a} di Padova, Padova Italy
\item \Idef{torino}INFN, Sezione di Torino, Turin, Italy
\item \Idef{det}Dipartimento DET del Politecnico di Torino, Turin, Italy
\item \Idef{disat}Dipartimento DISAT del Politecnico di Torino, Turin, Italy
\item \Idef{utorino}Dipartimento di Fisica dell Universit\`{a} di Torino, Turin, Italy
\item \Idef{piemonte}Dipartimento di Scienze e Innovazione Tecnologica dell'Universit\`{a} del Piemonte Orientale, Alessandria, Italy
\item \Idef{tohoku}Research Center for Electron Photon Science (ELPH), Tohoku University, Sendai, Japan
\item \Idef{cern}European Organization for Nuclear Research (CERN), Geneva, Switzerland
\item \Idef{rice}Department of Physics and Astronomy, Rice University, Houston, Texas, USA
\item \Idef{stonybrook}Department of Physics and Astronomy, Stony Brook University, SUNY, Stony Brook, New York, USA
\item \Idef{tamu}Cyclotron Institute and Department of Physics and Astronomy, Texas A\&M University, College Station, Texas, USA
\end{Authlist}
\endgroup